\documentclass[aps,prb,twocolumn,superscriptaddress,amsmath,amssymb,floatfix,longbibliography]{revtex4-2}
\usepackage{graphicx}
\usepackage{bm}
\usepackage{array}[=2016-10-06]
\usepackage{dcolumn}
\graphicspath{{./}}
\usepackage[hidelinks]{hyperref}

\newcommand{\EF}{E_{F}}
\newcommand{\kB}{k_{B}}

\newcommand{\mean}[1]{\left\langle #1 \right\rangle_{\theta}}
\newcommand{\dd}{\mathrm{d}}
\newcommand{\Li}{\mathrm{Li}}
\newcommand{\F}{\mathcal{F}}

\makeatletter
\newenvironment{appendixtable}
  {\onecolumngrid\@booleanfalse\floats@sw\begin{table*}\begin{minipage}{\textwidth}}
  {\end{minipage}\end{table*}\twocolumngrid}
\makeatother

\begin{document}

\title{Anisotropic Boltzmann transport in a tight-binding superlattice:\\
Exact results for power-law energy-dependent scattering}

\author{Sankar Das Sarma}
\author{Yuqing Lin}
\affiliation{Condensed Matter Theory Center and Joint Quantum Institute, Department of Physics, University of Maryland, College Park, Maryland 20742, USA}

\date{\today}

\begin{abstract}
We develop a complete Boltzmann transport theory for the in-plane ($\sigma_{xx}$) and out-of-plane ($\sigma_{zz}$) conductivities of an electronic superlattice consisting of a periodic stack of two-dimensional (2D) parabolic-band layers coupled along the superlattice axis by nearest-neighbor tunneling $t$, in the physically relevant regime $t\ll \EF$ where the Fermi surface is an open, corrugated cylinder. Within the relaxation-time approximation with a power-law energy-dependent scattering time $\tau(E)\propto E^{p}$, we study the seven models $p=0,\pm1,\pm3/2,\pm2$ as functions of the 2D electron density $n$ per layer and temperature $T$. For $p=-2$ (in plane) and $p\le -1$ (out of plane), long-lived carriers at the miniband bottom can dominate transport at finite temperature. For $p=-2$, these carriers produce a pronounced peak in $\sigma_{zz}(T)$ at $T\sim\EF$. We find that the temperature dependence of the anisotropic metallic dc conductivity can be highly counterintuitive depending on $t$, $T$, and $p$, with no generic increase or decrease with temperature. The Lorenz ratio is determined by the energy variance of the current-carrying electrons; the Wiedemann--Franz (WF) law is exact as $T\to0$ for all models at fixed $t$, with anisotropic corrections at finite temperature. Band-edge carriers first enhance the interlayer Lorenz ratio up to several times its low-temperature value through thermal occupation of separated energy ranges within the same electron band and then suppress it at high temperature by energy filtering. Strong violations of the Wiedemann--Franz law can therefore arise from miniband structure and energy-dependent elastic scattering alone.
\end{abstract}

\maketitle

\section{Introduction}
\label{sec:intro}

Electronic transport is central to solid-state physics: it distinguishes conductors from insulators and superconductors from non-superconductors, enables devices such as transistors, and provides signatures of physical phenomena such as quantum Hall effects, Majorana fermions, giant magnetoresistance, and strange metals. In fact, the term ``non-Fermi liquid'' is often applied to materials based just on the fact that their dc resistivity in a nominally metallic state does not follow the ``usual'' behavior, such as being proportional to $T^2$ at low temperatures or to $T$ at higher temperatures. The failure of the celebrated Wiedemann--Franz law, where the Lorenz number is a universal constant for metals, is often interpreted as evidence for ``non-Fermi-liquid'' physics~\cite{LavasaniBulmashDasSarma2019}. Understanding qualitative aspects of temperature-dependent electronic transport is arguably ``the'' central topic in condensed matter physics.

In this paper, we ask a simple conceptual question: given an anisotropic metal, where transport in one direction is constrained by a tight-binding model (with an effectively heavy mass and possibly an open Fermi surface)~\cite{Yamaji1989,McKenzieMoses1998}, what can one say about its dc transport properties based on the degree of anisotropy, Fermi energy, and various models for resistive scattering processes? The layered anisotropic metal model employed throughout is inspired by actual physical systems such as semiconductor superlattices~\cite{EsakiTsu1970,Bastard1988,Wacker2002}, multilayer graphene~\cite{McCannKoshino2013}, and even high-$T_c$ cuprates; our aim is to obtain general analytical insight into transport in layered metals, rather than to predict properties of any specific physical system. We find that transport even in this simple, idealized model of an anisotropic metal could manifest highly nontrivial and counterintuitive behavior depending on the details of the controlling Hamiltonian, leading to the conclusion that one should use extreme caution about jumping to conclusions regarding any possible exotic ground states based just on the qualitative behavior (e.g., temperature dependence) of the dc conductivity. Details matter very much even for the simplest possible models, and transport may manifest counterintuitive behavior arising from explicable physics without invoking any exotic possibilities.

The high-$T_c$ cuprates are also well-known examples of layered anisotropic ``metals,'' although they are much more complicated than semiconductor superlattices because the undoped system is most likely a Mott--Slater insulator~\cite{StepanovEtAl2026}.

The purpose of this paper is to provide a complete, essentially exact, and pedagogically transparent solution of the linear-response Boltzmann transport problem for this model in the relaxation-time approximation, for a family of power-law energy-dependent scattering times
\begin{equation}
\tau(E)=\tau_{0}\left(\frac{E}{E_{0}}\right)^{p},\qquad p\in\{0,\pm1,\pm\tfrac32,\pm2\},
\label{eq:tau}
\end{equation}
where $E$ is the total electron energy measured from the bottom of the miniband. Here $E_0$ is a fixed reference energy and $\tau_0=\tau(E_0)$. Power-law relaxation times are the standard phenomenological (and often microscopically justified) description of elastic and quasielastic scattering in semiconductors and metals~\cite{Ziman1960,AshcroftMermin,Lundstrom2000,GantmakherLevinson}: $p=0$ describes short-range (zero-range) disorder in 2D and acoustic-phonon scattering in the equipartition regime in 2D~\cite{Price1981,KawamuraDasSarma1992}; $p=1$ describes unscreened Coulomb impurity scattering in 2D~\cite{AndoFowlerStern1982,DasSarmaHwang2004}; $p=3/2$ describes ionized-impurity scattering in three dimensions (3D) up to logarithmic factors~\cite{ConwellWeisskopf1950,Brooks1955} as well as remote-impurity scattering in modulation-doped 2D systems with large setback~\cite{AndoFowlerStern1982}; negative exponents describe scattering channels whose phase space grows with energy, e.g., deformation-potential acoustic phonons in 3D ($p=-1/2$)~\cite{BardeenShockley1950}. We treat the seven values in Eq.~\eqref{eq:tau} on an equal footing: the positive values bracket the Coulomb-dominated regimes relevant to doped semiconductor structures, and the negative values are treated phenomenologically and bracket scattering rates that increase strongly with energy. In real 2D systems the temperature dependence of the resistivity at low $T$ is often dominated by the temperature dependence of screening~\cite{Stern1980,GoldDolgopolov1986,DasSarmaHwang1999}. This effect is excluded here by assuming a temperature-independent $\tau(E)$, so that the entire temperature dependence arises from Fermi--Dirac thermal averaging and from the temperature dependence of the chemical potential. The results obtained here are thus the ``kinematic'' baseline on which any microscopic temperature dependence of the scattering must be superposed. Appendix~\ref{app:scattering} summarizes standard scattering exponents and the assumptions underlying their microscopic assignments. In particular, phonon equipartition generally supplies an explicit temperature-dependent prefactor that is excluded from Eq.~\eqref{eq:tau}.

We work throughout in the regime $t\ll\EF$, where
$\EF=\pi\hbar^2n/m$ is the 2D Fermi energy of each layer.
This regime is natural for weakly coupled layers at moderate
densities and ensures an open Fermi surface.
Closed-form transport functions reduce both conductivities
exactly to one-dimensional energy integrals.
For the negative scattering exponents studied here,
long-lived carriers near the miniband bottom can produce
pronounced nonmonotonic interlayer transport as they become
thermally active. These carriers can also enhance the interlayer
Lorenz ratio before suppressing it at high temperature through
energy filtering. Miniband structure and energy-dependent
elastic scattering alone can thus produce strongly anisotropic
temperature dependences and substantial finite-temperature
violations of the Wiedemann--Franz law.

Sections~\ref{sec:model} and \ref{sec:boltzmann} introduce the model and Boltzmann transport formulation. Section~\ref{sec:electrical} develops the analytical results for electrical transport, including the zero-temperature, degenerate, and nondegenerate limits and band-edge contributions. Section~\ref{sec:numerics} presents the numerical results, with detailed validation in Appendix~\ref{app:validation}. Section~\ref{sec:WF} examines thermal transport and the Wiedemann--Franz law. Sections~\ref{sec:discussion} and \ref{sec:cuprates} discuss the physical implications, validity of the theory, and possible connections to layered cuprates, followed by the conclusions in Sec.~\ref{sec:conclusion}. Technical details are collected in the Appendixes.

\section{Model}
\label{sec:model}

\subsection{Band structure}

We consider an infinite periodic stack of identical 2D layers
in the $xy$ plane, spaced by $d$ along $z$.
In-plane motion is described by a parabolic dispersion with
effective mass $m$, and interlayer motion by nearest-neighbor
tight-binding hopping $t>0$. Retaining only the lowest
miniband, the dispersion is
\begin{equation}
E_{\bm k}=\frac{\hbar^{2}(k_x^{2}+k_y^{2})}{2m}
+2t\left(1-\cos k_zd\right),
\qquad |k_z|\le\frac{\pi}{d},
\label{eq:dispersion}
\end{equation}
with the energy zero at the band bottom. Expanding near
$k_z=0$ gives the longitudinal band mass
\begin{equation}
m_z=\frac{\hbar^{2}}{2td^{2}},
\qquad
\frac{m}{m_z}=\frac{2mtd^{2}}{\hbar^{2}}
=2\pi nd^{2}\,\frac{t}{\EF},
\label{eq:mz}
\end{equation}
where we used $\EF=\pi\hbar^{2}n/m$ (spin degeneracy 2).
The group velocities are
\begin{equation}
\bm v_{\parallel}=\frac{\hbar\bm k_{\parallel}}{m},
\qquad
v_z=\frac{2td}{\hbar}\sin k_zd.
\label{eq:velocity}
\end{equation}
Writing $\theta\equiv k_zd\in(-\pi,\pi]$,
$\varepsilon\equiv\hbar^{2}k_{\parallel}^{2}/2m$, and
$w(\theta)\equiv2t(1-\cos\theta)=4t\sin^{2}(\theta/2)$
gives $E=\varepsilon+w(\theta)$.
We denote the average over the Brillouin zone along $z$ by
$\mean{\cdots}=\int_{-\pi}^{\pi}(\dd\theta/2\pi)(\cdots)$.

\begin{figure*}[t]
\includegraphics[width=\textwidth]{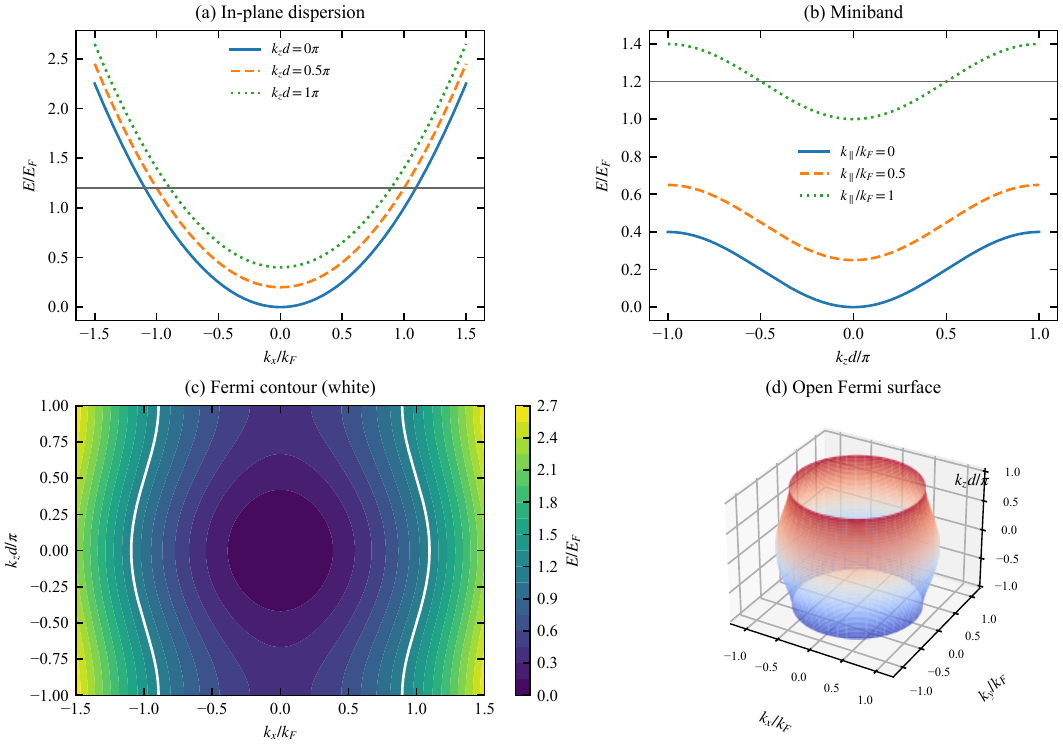}
\caption{Band structure of the superlattice, Eq.~\eqref{eq:dispersion}, drawn for $t=0.1\,\EF$ to make the $z$ dispersion visible (transport calculations use $t=0.01\,\EF$ unless otherwise stated). (a) In-plane dispersion at $k_{z}d=0,\pi/2,\pi$ (identical along $k_{x}$ and $k_{y}$). The horizontal line is the $T=0$ chemical potential $\mu_{0}=\EF+2t$. (b) Dispersion along $k_{z}$ at fixed $k_{\parallel}=0$, $0.5\,k_{F}$, $k_{F}$: a cosine miniband of width $4t$ and band-bottom mass $m_{z}=\hbar^{2}/2td^{2}$. (c) $E(k_{x},k_{y}=0,k_{z})$; the white line is the Fermi contour, which is open. (d) The three-dimensional Fermi surface, a corrugated cylinder.}
\label{fig:bands}
\end{figure*}

Figure~\ref{fig:bands} shows the band structure.
Along either in-plane direction [Fig.~\ref{fig:bands}(a)],
the dispersion is a parabola shifted by $w(\theta)$;
along $z$ [Fig.~\ref{fig:bands}(b)], it is a cosine miniband
of width $4t$. For $\EF>2t$, the Fermi surface is an open,
corrugated cylinder [Figs.~\ref{fig:bands}(c) and
\ref{fig:bands}(d)], with radius ranging from
$k_F\sqrt{1-2t/\EF}$ at $\theta=\pi$ to
$k_F\sqrt{1+2t/\EF}$ at $\theta=0$,
where $k_F=\sqrt{2\pi n}$.

\subsection{Density of states}

\begin{figure}[t]
\includegraphics[width=\columnwidth]{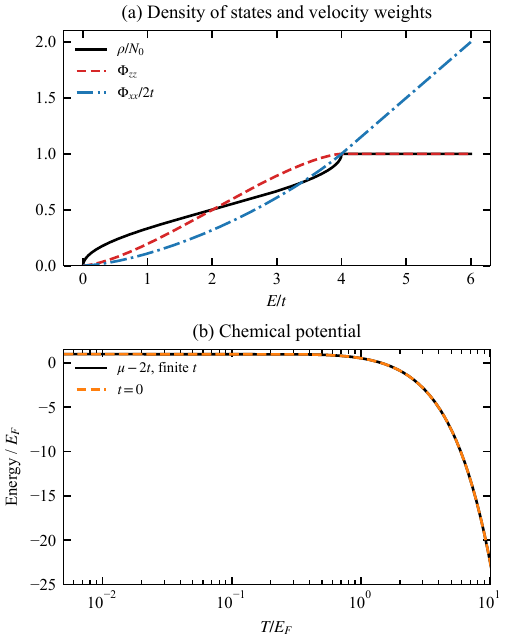}
\caption{(a) Density of states $\rho(E)/N_{0}$, Eq.~\eqref{eq:dos}, and the transport functions $\Phi_{xx}(E)/2t$ and $\Phi_{zz}(E)$, Eqs.~\eqref{eq:Phixx} and \eqref{eq:Phizz}, near the miniband. Above the miniband top, $E>4t$, the DOS and $\Phi_{zz}$ are constant, while $\Phi_{xx}$ is linear. (b) Chemical potential $\mu(T)-2t$ at fixed density for $t=0.01\,\EF$ (solid), compared with the strictly 2D result $T\ln(e^{\EF/T}-1)$ (dashed).}
\label{fig:dos}
\end{figure}

The density of states (DOS) per layer per unit area, including spin, is
\begin{equation}
\rho(E)=N_{0}\mean{\Theta\big(E-w(\theta)\big)}=
\begin{cases}
N_{0}\,\theta_{E}/\pi, & 0\le E\le4t,\\
N_{0}, & E>4t,
\end{cases}
\label{eq:dos}
\end{equation}
where $\Theta$ is the Heaviside step function, $N_{0}=m/\pi\hbar^{2}$ is the 2D DOS, and
\begin{equation}
\theta_{E}\equiv\arccos\left(1-\frac{E}{2t}\right)\in[0,\pi]
\label{eq:thetaE}
\end{equation}
is the value of $|k_{z}d|$ at which $w(\theta)=E$ for $0 \le E \le 4t$; for $E>4t$ we set $\theta_E=\pi$. Near the band bottom, $\rho(E)\simeq(N_{0}/\pi)(E/t)^{1/2}$, which is precisely $d$ times the 3D DOS of an anisotropic parabolic band with masses $(m,m,m_{z})$, as it must be. For $E>4t$ the DOS is exactly the 2D value. The DOS is shown in Fig.~\ref{fig:dos}(a).

\subsection{Chemical potential}

The electron density per layer, which is the same in every layer, is
\begin{equation}
n=\int_{0}^{\infty}\dd E\,\rho(E)f(E)=N_{0}\,\mean{T\ln\left(1+e^{[\mu-w(\theta)]/T}\right)},
\label{eq:density}
\end{equation}
where $f(E)=[e^{(E-\mu)/T}+1]^{-1}$ and we set $\kB=1$. The corresponding 3D density is $n_{3D}=n/d$. Since $n=N_{0}\EF$, Eq.~\eqref{eq:density} determines $\mu(n,T)$.

At $T=0$ and for $\mu_{0}>4t$,
$n=N_{0}\mean{\mu_{0}-w}=\mbox{$N_{0}(\mu_{0}-2t)$}$,
so that
\begin{equation}
\mu_{0}=\EF+2t\qquad\text{(exact for }\EF>2t\text{)}.
\label{eq:mu0}
\end{equation}
The condition $\mu_{0}>4t$, i.e., $\EF>2t$, is exactly the condition for the Fermi surface to be open. For $T\ll\EF$, writing $T\ln(1+e^{x})=Tx+T\ln(1+e^{-x})$, one finds
\begin{equation}
\mu=\EF+2t-T\,e^{-\EF/T}I_{0}\!\left(\frac{2t}{T}\right)+\cdots,
\label{eq:mulowT}
\end{equation}
where $I_{0}$ is a modified Bessel function; there is no power-law temperature dependence, exactly as for a strictly 2D parabolic band. In the nondegenerate regime,
\begin{equation}
e^{\mu/T}=\frac{\EF}{T}\,\frac{e^{2t/T}}{I_{0}(2t/T)}\left[1+O\!\left(\frac{\EF}{T}\right)\right],
\label{eq:muhighT}
\end{equation}
and in the limit $t\to0$ one recovers the exact 2D result,
\begin{equation}
\mu=T\ln\left(e^{\EF/T}-1\right),\qquad z\equiv e^{\mu/T}=e^{\EF/T}-1,
\label{eq:mu2D}
\end{equation}
which defines the fugacity $z$ used below. The numerically computed $\mu(T)-2t$ is compared with Eq.~\eqref{eq:mu2D} in Fig.~\ref{fig:dos}(b).

\section{Boltzmann theory}
\label{sec:boltzmann}

\subsection{Conductivity tensor}

In the linearized Boltzmann equation with a relaxation time that depends only on energy, the nonequilibrium distribution in a weak uniform electric field $\bm{\mathcal E}$ is $f_{\bm k}=f(E_{\bm k})+e\tau(E_{\bm k})\,\bm v_{\bm k}\cdot\bm{\mathcal E}\,\partial f/\partial E$ (electron charge $-e$), and the conductivity tensor is~\cite{Ziman1960,AshcroftMermin}
\begin{equation}
\sigma_{ij}=2e^{2}\int\frac{\dd^{3}k}{(2\pi)^{3}}\,\tau(E_{\bm k})\,v_{i}v_{j}\left(-\frac{\partial f}{\partial E}\right)_{E=E_{\bm k}} .
\label{eq:sigmaij}
\end{equation}
By symmetry $\sigma_{ij}$ is diagonal with $\sigma_{xx}=\sigma_{yy}\neq\sigma_{zz}$. Let $\varphi$ denote the in-plane azimuthal angle. Using $\int\dd^{3}k/(2\pi)^{3}=d^{-1}\int\dd^{2}k_{\parallel}/(2\pi)^{2}\int_{-\pi}^{\pi}\dd\theta/2\pi$, the in-plane angular average $\langle v_{x}^{2}\rangle_{\varphi}=v_{\parallel}^{2}/2=\varepsilon/m$, and Eq.~\eqref{eq:velocity}, \mbox{we obtain}
\begin{align}
\sigma_{xx}&=\frac{e^{2}N_{0}}{md}\,\mean{\int_{0}^{\infty}\dd\varepsilon\,\varepsilon\,\tau(\varepsilon+w)\left(-\frac{\partial f}{\partial E}\right)_{\varepsilon+w}},
\label{eq:sxx2d}\\
\sigma_{zz}&=\frac{4e^{2}N_{0}t^{2}d}{\hbar^{2}}\nonumber\\
&\quad\times\mean{\sin^{2}\theta\int_{0}^{\infty}\dd\varepsilon\,\tau(\varepsilon+w)\left(-\frac{\partial f}{\partial E}\right)_{\varepsilon+w}}.
\label{eq:szz2d}
\end{align}

\subsection{Transport functions}

Changing variables from $\varepsilon$ to $E=\varepsilon+w$ and performing the $\theta$ average at fixed $E$ reduces Eqs.~\eqref{eq:sxx2d} and \eqref{eq:szz2d} exactly to one-dimensional integrals,
\begin{align}
\sigma_{xx}&=\frac{e^{2}N_{0}}{md}\int_{0}^{\infty}\dd E\,\tau(E)\,\Phi_{xx}(E)\left(-\frac{\partial f}{\partial E}\right),
\label{eq:sxx1d}\\
\sigma_{zz}&=\frac{2e^{2}N_{0}t^{2}d}{\hbar^{2}}\int_{0}^{\infty}\dd E\,\tau(E)\,\Phi_{zz}(E)\left(-\frac{\partial f}{\partial E}\right),
\label{eq:szz1d}
\end{align}
with the transport functions
\begin{align}
\Phi_{xx}(E)&\equiv\mean{(E-w)\,\Theta(E-w)}\nonumber\\
&=\frac{1}{\pi}\Big[(E-2t)\,\theta_{E}+2t\sin\theta_{E}\Big],
\label{eq:Phixx}\\
\Phi_{zz}(E)&\equiv2\mean{\sin^{2}\theta\;\Theta(E-w)}\nonumber\\
&=\frac{1}{\pi}\Big[\theta_{E}-\sin\theta_{E}\cos\theta_{E}\Big],
\label{eq:Phizz}
\end{align}
where $\theta_{E}$ is defined in Eq.~\eqref{eq:thetaE} and $\theta_{E}=\pi$ for $E\ge4t$. Above the miniband top the transport functions are elementary,
\begin{equation}
\Phi_{xx}(E)=E-2t,\qquad \Phi_{zz}(E)=1\qquad(E\ge4t),
\label{eq:Phiabove}
\end{equation}
while near the band bottom
\begin{equation}
\Phi_{xx}(E)\simeq\frac{2}{3\pi}\frac{E^{3/2}}{t^{1/2}},\quad \Phi_{zz}(E)\simeq\frac{2}{3\pi}\left(\frac{E}{t}\right)^{3/2}\quad(E\ll t),
\label{eq:Phibottom}
\end{equation}
which is the behavior of $v_{i}^{2}$ times the DOS of an anisotropic 3D parabolic band. Physically, $\Phi_{xx}$ and $\Phi_{zz}$ are the energy-resolved ``transport densities of states'' $\propto\sum_{\bm k}v_{i}^{2}\delta(E-E_{\bm k})$, shown in Fig.~\ref{fig:dos}(a). All results below follow from Eqs.~\eqref{eq:sxx1d}--\eqref{eq:Phibottom}. For finite $t,T>0$, both band-bottom integrals converge if and only if $p>-5/2$, which includes all exponents considered here.

\subsection{Exact sum rule and Drude form}

Since $\dd\Phi_{xx}/\dd E=\mean{\Theta(E-w)}=\rho(E)/N_{0}$, an integration by parts gives, for arbitrary $T$ and $t$,
\begin{equation}
\int_{0}^{\infty}\dd E\,\Phi_{xx}(E)\left(-\frac{\partial f}{\partial E}\right)=\frac{1}{N_{0}}\int_{0}^{\infty}\dd E\,\rho(E)f(E)=\EF .
\label{eq:sumrule}
\end{equation}
Consequently the in-plane conductivity has \emph{exactly} the Drude form
\begin{equation}
\sigma_{xx}=\frac{n_{3D}e^{2}}{m}\,\langle\!\langle\tau\rangle\!\rangle,\qquad
\langle\!\langle\tau\rangle\!\rangle\equiv\frac{\int\dd E\,\tau\,\Phi_{xx}(-\partial f/\partial E)}{\int\dd E\,\Phi_{xx}(-\partial f/\partial E)},
\label{eq:drude}
\end{equation}
with the full 3D density and a transport-averaged relaxation time. An immediate corollary is that for energy-independent scattering ($p=0$),
\begin{equation}
\sigma_{xx}=\frac{ne^{2}\tau_{0}}{md}\qquad\text{exactly, for all }T\text{ and }t,
\label{eq:xxp0exact}
\end{equation}
which the numerics reproduce to machine precision. No such sum rule holds for $\sigma_{zz}$, since $\Phi_{zz}$ weights the states by $v_{z}^{2}$, which is not proportional to the kinetic energy.

\subsection{Conductivity normalization}

For the temperature dependence at fixed density we measure $\tau$ in units of $\tau(\EF)$ and use the reference conductivities
\begin{align}
\sigma_{xx}^{0}&=\frac{ne^{2}\tau(\EF)}{md}=\frac{e^{2}N_{0}}{md}\,\EF\,\tau(\EF),
\label{eq:s0xx}\\
\sigma_{zz}^{0}&=\frac{2e^{2}N_{0}t^{2}d}{\hbar^{2}}\,\tau(\EF)=\frac{e^{2}N_{0}\,t\,\tau(\EF)}{m_{z}d}.
\label{eq:s0zz}
\end{align}
Their ratio is
\begin{equation}
\frac{\sigma_{zz}^{0}}{\sigma_{xx}^{0}}=\frac{m}{m_{z}}\,\frac{t}{\EF}=2\pi nd^{2}\left(\frac{t}{\EF}\right)^{2}.
\label{eq:ratio0}
\end{equation}
For the density dependence at fixed temperature we fix $E_{0}=\EF(n_{0})$ for a reference density $n_{0}$, and measure conductivities in units of $n_{0}e^{2}\tau_{0}/md$ and $2e^{2}mt^{2}d\,\tau_{0}/\pi\hbar^{4}$.

\section{Electrical conductivity}
\label{sec:electrical}

\subsection{Zero temperature}
\label{sec:T0}

At $T=0$, $-\partial f/\partial E=\delta(E-\mu_{0})$ and, since $\mu_{0}=\EF+2t>4t$, Eqs.~\eqref{eq:Phiabove} and \eqref{eq:sumrule} give the results
\begin{align}
\sigma_{xx}(T=0)&=\frac{ne^{2}}{md}\,\tau(\EF+2t),
\label{eq:sxxT0}\\
\sigma_{zz}(T=0)&=\frac{2e^{2}N_{0}t^{2}d}{\hbar^{2}}\,\tau(\EF+2t),
\label{eq:szzT0}
\end{align}
which are exact for any $t<\EF/2$ and any $\tau(E)$.
For the power-law scattering time of Eq.~\eqref{eq:tau},
\begin{equation}
\frac{\sigma_{xx}(0)}{\sigma_{xx}^{0}}=\frac{\sigma_{zz}(0)}{\sigma_{zz}^{0}}=\left(1+\frac{2t}{\EF}\right)^{p}.
\label{eq:T0power}
\end{equation}
At fixed $t$, $E_0$, and $\tau_0$, the leading density
dependences are $\sigma_{xx}(0)\propto n^{1+p}$ and
$\sigma_{zz}(0)\propto n^{p}$ for $t\ll\EF$.
The interlayer result reflects the constant 2D DOS:
on an open Fermi surface, the angular weight is uniform,
giving $\mean{v_z^2}=2t^2d^2/\hbar^2$.
Equivalently, one may write
$\sigma_{zz}(0)=n_{3D}e^2\tau/m_{zz}^{*}$ with
$m_{zz}^{*}=m_z\EF/t\gg m_z$: electrons on an open
Fermi surface do not experience the band-bottom mass $m_z$.

\subsection{Degenerate regime: Sommerfeld expansion}
\label{sec:degenerate}

For $T\ll\EF$, the integrands of Eqs.~\eqref{eq:sxx1d} and
\eqref{eq:szz1d} near $E=\mu$ are
$G_{xx}(E)=\tau(E)(E-2t)$ and $G_{zz}(E)=\tau(E)$, which
are analytic there, because $\Phi_{xx}$ and $\Phi_{zz}$
are polynomials for $E>4t$. The Sommerfeld
expansion~\cite{AshcroftMermin}
\begin{align}
\int\dd E\,G(E)\left(-\frac{\partial f}{\partial E}\right)
&=G(\mu)+\frac{\pi^{2}}{6}T^{2}G''(\mu)\nonumber\\
&\quad+\frac{7\pi^{4}}{360}T^{4}G''''(\mu)+\cdots
\label{eq:sommerfeld}
\end{align}
is then valid up to exponentially small terms, and
$\mu=\mu_{0}$ up to exponentially small terms
[Eq.~\eqref{eq:mulowT}]. Evaluating the derivatives exactly
in $t$ we obtain
\begin{widetext}
\begin{align}
\frac{\sigma_{xx}(T)}{\sigma_{xx}(0)}
&=1+\frac{\pi^{2}}{6}\,p
\left[(p-1)+\frac{2\mu_{0}}{\EF}\right]
\left(\frac{T}{\mu_{0}}\right)^{2}
+\frac{7\pi^{4}}{360}\,p(p-1)(p-2)
\left[(p-3)+\frac{4\mu_{0}}{\EF}\right]
\left(\frac{T}{\mu_{0}}\right)^{4}
+O\!\left(T^{6}\right),
\label{eq:sommxx}\\
\frac{\sigma_{zz}(T)}{\sigma_{zz}(0)}
&=1+\frac{\pi^{2}}{6}\,p(p-1)
\left(\frac{T}{\mu_{0}}\right)^{2}
+\frac{7\pi^{4}}{360}\,p(p-1)(p-2)(p-3)
\left(\frac{T}{\mu_{0}}\right)^{4}
+O\!\left(T^{6}\right).
\label{eq:sommzz}
\end{align}
\noindent
Here $\mu_{0}=\EF+2t$, and $\sigma(0)$ is given by
Eqs.~\eqref{eq:sxxT0} and \eqref{eq:szzT0}.
To leading order in $t/\EF$,
\begin{align}
\frac{\sigma_{xx}}{\sigma_{xx}^{0}}
&\simeq\left(1+\frac{2pt}{\EF}\right)
\left[
1+\frac{\pi^{2}}{6}p(p+1)\frac{T^{2}}{\EF^{2}}
+\frac{7\pi^{4}}{360}(p+1)p(p-1)(p-2)
\frac{T^{4}}{\EF^{4}}
\right],
\label{eq:sommxx0}\\
\frac{\sigma_{zz}}{\sigma_{zz}^{0}}
&\simeq\left(1+\frac{2pt}{\EF}\right)
\left[
1+\frac{\pi^{2}}{6}p(p-1)\frac{T^{2}}{\EF^{2}}
+\frac{7\pi^{4}}{360}p(p-1)(p-2)(p-3)
\frac{T^{4}}{\EF^{4}}
\right].
\label{eq:sommzz0}
\end{align}
\end{widetext}
The coefficient of $T^{2}$ in $\sigma_{xx}$, $p(p+1)$,
vanishes for $p=0$ and $p=-1$; in $\sigma_{zz}$,
$p(p-1)$ vanishes for $p=0$ and $p=1$. For these four
cases all power-law corrections vanish (to leading order
in $t$) and the temperature dependence is exponentially
small (Sec.~\ref{sec:exact}). For all other $p$ in our set
the $T^{2}$ coefficient is positive, so thermal smearing
\emph{increases} the conductivity at low $T$: for $p>0$
because higher-energy electrons have longer lifetimes;
for $p<-1$ ($\sigma_{xx}$) and $p<0$ ($\sigma_{zz}$)
because $\tau(E)$ is convex. The Sommerfeld expansion
misses one additional, exponentially small but
parametrically large, contribution from the band bottom
for the singular exponents; this is treated in
Sec.~\ref{sec:edge}.

\subsection{Exact solution to leading order in $t/\EF$ for all temperatures}
\label{sec:exact}

\subsubsection[Fermi--Dirac representation and special cases]{Fermi--Dirac representation and special cases}

For $t\to0$ at fixed $T$ we may set $\Phi_{xx}=E$ and $\Phi_{zz}=1$ in Eqs.~\eqref{eq:sxx1d} and \eqref{eq:szz1d}, and $\mu$ is given by Eq.~\eqref{eq:mu2D}, provided the resulting integrals converge at $E\to0$. The required integral follows from the expansion $-\partial f/\partial E=T^{-1}\sum_{k\ge1}(-1)^{k+1}k\,z^{k}e^{-kE/T}$ (for $z<1$, and by analytic continuation for all $z>0$):
\begin{equation}
\int_{0}^{\infty}\dd E\,E^{a}\left(-\frac{\partial f}{\partial E}\right)=\Gamma(a+1)\,T^{a}\,\F_{a}(z),\qquad a>-1,
\label{eq:FDmoment}
\end{equation}
where
\begin{equation}
\F_{s}(z)\equiv-\Li_{s}(-z)=\frac{1}{\Gamma(s)}\int_{0}^{\infty}\frac{x^{s-1}\,\dd x}{z^{-1}e^{x}+1}
\label{eq:FDint}
\end{equation}
is the complete Fermi--Dirac integral (the second form holding for $s>0$), $\Gamma$ is the Euler gamma function, and $\Li_{s}$ is the polylogarithm~\cite{DLMF,Blakemore1982}. For all $T/\EF$, we~obtain
\begin{align}
\frac{\sigma_{xx}}{\sigma_{xx}^{0}}&=\Gamma(p+2)\left(\frac{T}{\EF}\right)^{p+1}\F_{p+1}\!\left(e^{\EF/T}-1\right),
\label{eq:exactxx}\\
\frac{\sigma_{zz}}{\sigma_{zz}^{0}}&=\Gamma(p+1)\left(\frac{T}{\EF}\right)^{p}\F_{p}\!\left(e^{\EF/T}-1\right),
\label{eq:exactzz}
\end{align}
valid for $p>-2$ and $p>-1$, respectively, with corrections of relative order $t/\EF$ in the degenerate Sommerfeld expansion [where they are given by Eqs.~\eqref{eq:sommxx} and \eqref{eq:sommzz}] and $t/T$ (up to logarithms) or $(t/T)^{1/2}$ in the nondegenerate regime (Sec.~\ref{sec:nondegenerate}). Comparing Eqs.~\eqref{eq:exactxx} and \eqref{eq:exactzz} reveals the identity
\begin{equation}
\frac{\sigma_{xx}(p)}{\sigma_{xx}^{0}(p)}=\frac{\sigma_{zz}(p+1)}{\sigma_{zz}^{0}(p+1)}\qquad(t\to0),
\label{eq:identity}
\end{equation}
because the in-plane velocity weight $v_x^2\propto\varepsilon$
adds one power of energy, while the averaged interlayer
weight $v_z^2$ is energy independent.

Using $\F_{1}(z)=\ln(1+z)$, $\F_{0}(z)=z/(1+z)$, and $\F_{-1}(z)=z/(1+z)^{2}$, the cases in our set are, explicitly:

\noindent\emph{In-plane conductivity:}
\begin{align}
p=0:&\quad \sigma_{xx}=\sigma_{xx}^{0}\quad\text{for all }T\text{ (exact for all }t\text{)},
\label{eq:xxp0}\\
p=-1:&\quad \sigma_{xx}=\sigma_{xx}^{0}\left(1-e^{-\EF/T}\right),
\label{eq:xxpm1}\\
p=1:&\quad \sigma_{xx}=2\sigma_{xx}^{0}\left(\frac{T}{\EF}\right)^{2}\F_{2}\!\left(e^{\EF/T}-1\right),
\label{eq:xxp1}\\
p=2:&\quad \sigma_{xx}=6\sigma_{xx}^{0}\left(\frac{T}{\EF}\right)^{3}\F_{3}\!\left(e^{\EF/T}-1\right),
\label{eq:xxp2}\\
p=\tfrac32:&\quad \sigma_{xx}=\frac{15\sqrt\pi}{8}\sigma_{xx}^{0}\left(\frac{T}{\EF}\right)^{5/2}\F_{5/2}\!\left(e^{\EF/T}-1\right),
\label{eq:xxp32}\\
p=-\tfrac32:&\quad \sigma_{xx}=\sqrt\pi\sigma_{xx}^{0}\left(\frac{T}{\EF}\right)^{-1/2}\F_{-1/2}\!\left(e^{\EF/T}-1\right).
\label{eq:xxpm32}
\end{align}
\emph{Interlayer conductivity:}
\begin{align}
p=1:&\quad \sigma_{zz}=\sigma_{zz}^{0}\quad\text{exactly, for all }T\text{ as }t\to0,
\label{eq:zzp1}\\
p=0:&\quad \sigma_{zz}=\sigma_{zz}^{0}\left(1-e^{-\EF/T}\right),
\label{eq:zzp0}\\
p=2:&\quad \sigma_{zz}=2\sigma_{zz}^{0}\left(\frac{T}{\EF}\right)^{2}\F_{2}\!\left(e^{\EF/T}-1\right),
\label{eq:zzp2}\\
p=\tfrac32:&\quad \sigma_{zz}=\frac{3\sqrt\pi}{4}\sigma_{zz}^{0}\left(\frac{T}{\EF}\right)^{3/2}\F_{3/2}\!\left(e^{\EF/T}-1\right).
\label{eq:zzp32}
\end{align}
For the remaining cases, $p=-2$ in $\sigma_{xx}$ and
$p=-1,-3/2,-2$ in $\sigma_{zz}$, the $t=0$ transport kernels
give divergent energy integrals at any $T>0$.
Near $E=0$, the integrands behave as $E^{p+1}$ ($\sigma_{xx}$)
or $E^p$ ($\sigma_{zz}$), since $-\partial f/\partial E$ is
finite. The miniband regularizes these divergent integrals by
suppressing the transport functions at $E\lesssim4t$,
Eq.~\eqref{eq:Phibottom}. The regularized results appear in
Secs.~\ref{sec:nondegenerate} and \ref{sec:edge}.

Equations~\eqref{eq:xxp0} and \eqref{eq:zzp1} highlight
temperature-independent transport at fixed density:
the in-plane conductivity for $p=0$ and, to leading order
in $t/\EF$, the interlayer conductivity for $p=1$ are
constant in both statistical regimes.
The in-plane result is exact for any $t$ by the sum rule,
Eq.~\eqref{eq:xxp0exact}. For interlayer $p=1$, finite-$t$
corrections are $O(t/\EF)$ in the degenerate regime, while
$\sigma_{zz}/\sigma_{zz}^{0}\simeq1+2t/T$ in the
nondegenerate regime.

\subsubsection{Complete degenerate series}

For $\eta\equiv\mu/T\to+\infty$ the Fermi--Dirac integral has the asymptotic expansion~\cite{AshcroftMermin,Blakemore1982}
\begin{equation}
\F_s(e^\eta)=\sum_{k=0}^{N}\frac{a_k\eta^{s-2k}}{\Gamma(s+1-2k)}
+O(\eta^{s-2N-2}),\qquad \eta\to\infty ,
\label{eq:Fasymp}
\end{equation}
with
\begin{equation}
a_{k}=2\left(1-2^{1-2k}\right)\zeta(2k),
\label{eq:ak}
\end{equation}
i.e., $a_{0}=1$, $a_{1}=\pi^{2}/6$, $a_{2}=7\pi^{4}/360$, $a_{3}=31\pi^{6}/15120$,
where $\zeta$ is the Riemann zeta function. In the degenerate regime $\eta=\ln(e^{\EF/T}-1)=\EF/T-e^{-\EF/T}+\cdots$, so that, to leading order in $t/\EF$,
\begin{align}
\frac{\sigma_{xx}}{\sigma_{xx}^{0}}&\simeq\sum_{k\ge0}a_{k}\,(p+1)^{\underline{2k}}\left(\frac{T}{\EF}\right)^{2k},
\label{eq:fullsommxx}\\
\frac{\sigma_{zz}}{\sigma_{zz}^{0}}&\simeq\sum_{k\ge0}a_{k}\,p^{\underline{2k}}\left(\frac{T}{\EF}\right)^{2k}.
\label{eq:fullsommzz}
\end{align}
Here $x^{\underline n}=x(x-1)\cdots(x-n+1)$ is the falling
factorial, with $x^{\underline0}=1$. These series follow
from Eq.~\eqref{eq:sommerfeld} carried to all orders.
For the singular cases, the expansion is performed at finite
$t$ before retaining the leading terms in $t/\EF$.
Coefficients exact in $t$ follow by replacing
$(p+1)^{\underline{2k}}\EF^{-2k}$ and
$p^{\underline{2k}}\EF^{-2k}$ with
$G_{xx}^{(2k)}(\mu_0)/G_{xx}(\mu_0)$ and
$G_{zz}^{(2k)}(\mu_0)/G_{zz}(\mu_0)$, respectively
(Appendix~\ref{app:sommerfeld}). The resulting series give
$\sigma_{ii}(T)/\sigma_{ii}(0)$; multiplication by
$(1+2t/\EF)^p$ restores the normalization to
$\sigma_{ii}^{0}$.
The leading-order series terminate when $p+1$ ($\sigma_{xx}$)
or $p$ ($\sigma_{zz}$) is a nonnegative integer;
otherwise they are asymptotic.
Hence, up to terms of order $e^{-\EF/T}$ and $t/\EF$,
\begin{align}
\sigma_{xx}(p=1)&=\sigma_{xx}^{0}\left[1+\frac{\pi^{2}}{3}\frac{T^{2}}{\EF^{2}}\right],
\label{eq:termxx1}\\
\sigma_{xx}(p=2)&=\sigma_{xx}^{0}\left[1+\pi^{2}\frac{T^{2}}{\EF^{2}}\right],
\label{eq:termxx2}\\
\sigma_{zz}(p=2)&=\sigma_{zz}^{0}\left[1+\frac{\pi^{2}}{3}\frac{T^{2}}{\EF^{2}}\right],
\label{eq:termzz2}
\end{align}
with no $T^{4}$ or higher power-law terms. The exponentially small remainders follow from the inversion formulas $\F_{2}(e^{\eta})=\eta^{2}/2+\pi^{2}/6-\F_{2}(e^{-\eta})$ and $\F_{3}(e^{\eta})=\eta^{3}/6+\pi^{2}\eta/6+\F_{3}(e^{-\eta})$~\cite{DLMF}, which give closed forms valid for all $T$:
\begin{widetext}
\begin{align}
\frac{\sigma_{xx}(p=1)}{\sigma_{xx}^{0}}&=\frac{\sigma_{zz}(p=2)}{\sigma_{zz}^{0}}=\left(\frac{\mu}{\EF}\right)^{2}+\frac{\pi^{2}}{3}\frac{T^{2}}{\EF^{2}}-2\frac{T^{2}}{\EF^{2}}\F_{2}\!\left(e^{-\mu/T}\right),
\label{eq:invp1}\\
\frac{\sigma_{xx}(p=2)}{\sigma_{xx}^{0}}&=\left(\frac{\mu}{\EF}\right)^{3}+\pi^{2}\frac{\mu}{\EF}\frac{T^{2}}{\EF^{2}}+6\frac{T^{3}}{\EF^{3}}\F_{3}\!\left(e^{-\mu/T}\right),
\label{eq:invp2}
\end{align}
\end{widetext}
with $\mu=T\ln(e^{\EF/T}-1)$ from Eq.~\eqref{eq:mu2D}. In the degenerate limit $\mu/\EF=1-(T/\EF)e^{-\EF/T}+\cdots$ and $\F_{s}(e^{-\mu/T})\simeq e^{-\EF/T}$, so that the leading exponential corrections are
\begin{align}
\delta\sigma_{xx}(p=1)/\sigma_{xx}^{0}&\simeq-2\frac{T}{\EF}\left(1+\frac{T}{\EF}\right)e^{-\EF/T},
\label{eq:expxx1}\\
\delta\sigma_{xx}(p=2)/\sigma_{xx}^{0}&\simeq-3\frac{T}{\EF}e^{-\EF/T}\nonumber\\
&\quad\times\left[1+\left(\frac{\pi^2}{3}-2\right)\frac{T^2}{\EF^2}\right].
\label{eq:expxx2}
\end{align}
Similarly, the elementary results \eqref{eq:xxpm1} and \eqref{eq:zzp0} show that for $\sigma_{xx}$ with $p=-1$ and $\sigma_{zz}$ with $p=0$ the \emph{entire} temperature dependence is contained in the factor $1-e^{-\EF/T}$, whose deviation from unity is exponentially small in the degenerate regime.
 
With $\vartheta\equiv T/\EF$, the half-integer cases give
\begin{align}
\frac{\sigma_{xx}(\tfrac32)}{\sigma_{xx}^{0}}&\simeq1+\frac{5\pi^{2}}{8}\vartheta^{2}-\frac{7\pi^{4}}{384}\vartheta^{4}-\frac{155\pi^{6}}{21504}\vartheta^{6},
\label{eq:hxx32}\\
\frac{\sigma_{xx}(-\tfrac32)}{\sigma_{xx}^{0}}&\simeq1+\frac{\pi^{2}}{8}\vartheta^{2}+\frac{49\pi^{4}}{384}\vartheta^{4}+\frac{341\pi^{6}}{1024}\vartheta^{6},
\label{eq:hxxm32}\\
\frac{\sigma_{zz}(\tfrac32)}{\sigma_{zz}^{0}}&\simeq1+\frac{\pi^{2}}{8}\vartheta^{2}+\frac{7\pi^{4}}{640}\vartheta^{4}+\frac{31\pi^{6}}{3072}\vartheta^{6},
\label{eq:hzz32}\\
\frac{\sigma_{zz}(-\tfrac32)}{\sigma_{zz}^{0}}&\simeq1+\frac{5\pi^{2}}{8}\vartheta^{2}+\frac{147\pi^{4}}{128}\vartheta^{4}+\frac{4433\pi^{6}}{1024}\vartheta^{6},
\label{eq:hzzm32}
\end{align}
and for the negative integers
\begin{align}
\frac{\sigma_{xx}(-2)}{\sigma_{xx}^{0}}&\simeq\frac{\sigma_{zz}(-1)}{\sigma_{zz}^{0}}\simeq1+\frac{\pi^{2}}{3}\vartheta^{2}+\frac{7\pi^{4}}{15}\vartheta^{4}+\frac{31\pi^{6}}{21}\vartheta^{6},
\label{eq:hxxm2}\\
\frac{\sigma_{zz}(-2)}{\sigma_{zz}^{0}}&\simeq1+\pi^{2}\vartheta^{2}+\frac{7\pi^{4}}{3}\vartheta^{4}+\frac{31\pi^{6}}{3}\vartheta^{6}.
\label{eq:hzzm2}
\end{align}
These displayed series retain order $(t/\EF)^0$. The finite-$t$ Sommerfeld coefficients are given in Eqs.~\eqref{eq:sommxx} and \eqref{eq:sommzz} and Appendix~\ref{app:sommerfeld}. For the singular cases, Eqs.~\eqref{eq:hxxm2} and \eqref{eq:hzzm2} [and \eqref{eq:hzzm32}] must be supplemented by the band-edge terms of Sec.~\ref{sec:edge}. The rapid growth of the coefficients for negative $p$ signals the breakdown of the degenerate expansion at $T\approx0.1$--$0.2\,\EF$, precisely where the band-edge term takes over.

\subsection{Nondegenerate regime}
\label{sec:nondegenerate}

For $T\gg\EF$ (and hence $T\gg t$) the distribution is Boltzmann-like, $-\partial f/\partial E\simeq(z/T)e^{-E/T}$ with $z\simeq\EF/T$, and the conductivities take the form
\begin{align}
\sigma_{xx}&=\frac{ne^{2}}{md}\,\tau(T)\,C_{xx}(p)\left[1+c_{1}\frac{\EF}{T}+c_{2}\frac{\EF^{2}}{T^{2}}+\cdots\right],
\label{eq:highTxx}\\
\sigma_{zz}&=\frac{2ne^{2}t^{2}d}{\hbar^{2}T}\,\tau(T)\,C_{zz}(p)\left[1+c_{1}\frac{\EF}{T}+c_{2}\frac{\EF^{2}}{T^{2}}+\cdots\right],
\label{eq:highTzz}
\end{align}
where $\tau(T)\equiv\tau_{0}(T/E_{0})^{p}$, the coefficients $C(p)$ have the small-$t/T$ expansions specified below, and $c_{1,2}$ (different for each component and each $p$) are the quantum-degeneracy corrections. In terms of the reference conductivities of Eqs.~\eqref{eq:s0xx} and \eqref{eq:s0zz}, $\sigma_{xx}/\sigma_{xx}^{0}=(T/\EF)^{p}C_{xx}[\cdots]$ and $\sigma_{zz}/\sigma_{zz}^{0}=(T/\EF)^{p-1}C_{zz}[\cdots]$. Equation~\eqref{eq:highTzz} has a transparent classical meaning: for a uniformly occupied miniband $\langle v_{z}^{2}\rangle=2t^{2}d^{2}/\hbar^{2}$, and the classical result $\sigma_{zz}=n_{3D}e^{2}\langle v_{z}^{2}\tau\rangle/T$ follows.

\subsubsection{Nonsingular cases}

For the nonsingular cases, Eq.~\eqref{eq:FDmoment} gives, to zeroth order in $t/T$,
\begin{equation}
C_{xx}=\Gamma(p+2),\qquad C_{zz}=\Gamma(p+1).
\label{eq:Cregular}
\end{equation}
At leading order in $t/T$, the small-$z$ expansion $\F_{s}(z)=\sum_{k\ge1}(-1)^{k+1}z^{k}k^{-s}$ combined with $z=e^{\EF/T}-1=x+x^{2}/2+x^{3}/6+\cdots$, $x\equiv\EF/T$, gives the quantum corrections
\begin{equation}
c_{1}=\frac12-2^{-s},\qquad c_{2}=\frac16-2^{-s}+3^{-s},
\label{eq:cquantum}
\end{equation}
with $s=p+1$ for $\sigma_{xx}$ and $s=p$ for $\sigma_{zz}$. Note that $c_{1}=c_{2}=0$ for $\sigma_{xx}(p=0)$ and $\sigma_{zz}(p=1)$, consistent with Eqs.~\eqref{eq:xxp0} and \eqref{eq:zzp1}, and that for $\sigma_{xx}(p=-1)$ and $\sigma_{zz}(p=0)$ one recovers $1-e^{-x}\simeq x(1-x/2+x^{2}/6)$.

The classical miniband corrections (Appendix~\ref{app:tcorr}), with $b=t/T\to0$, are
\begin{align}
C_{xx}(p)&=\Gamma(p+2)\left[1+\frac{2t}{T}\,\frac{p}{p+1}\right],\qquad p>-1,
\label{eq:tcorrxx}\\
C_{xx}(-1)&=1-\frac{2t}{T}\left[\ln\frac{T}{t}-\gamma-1\right],
\label{eq:tcorrxxm1}\\
C_{xx}(-\tfrac32)&=\sqrt\pi-\frac{16}{\pi}\sqrt{b}+6\sqrt\pi\,b+O(b^{3/2}),
\label{eq:Cxxm32}\\
C_{zz}(p)&=\Gamma(p+1)\left[1+\frac{2t}{T}\right],\qquad p>0,
\label{eq:tcorrzz}\\
C_{zz}(0)&=1+O\!\left(t^{2}/T^{2}\right),
\label{eq:tcorrzz0}
\end{align}
where $\gamma=0.5772\ldots$ is Euler's constant. The classical and quantum factors are separate asymptotic approximations; their product does not determine all mixed coefficients in $\EF/T$ and $t/T$. The correction for $\sigma_{xx}$ with $p=-3/2$ is anomalously large, $O[(t/T)^{1/2}]$, because the integrand $\varepsilon^{-1/2}$ is dominated by low energies, and that for $p=-1$ carries a logarithm. 
\subsubsection{Singular cases}

For the singular cases the $\theta$ averages must be performed with the full $w(\theta)$; the necessary integrals are listed in Appendix~\ref{app:integrals}. We find
\begin{align}
C_{xx}(-2)&=L_{xx}+\frac{4t}{T}\left[\ln\frac{T}{t}-\gamma\right],\quad L_{xx}\equiv\ln\frac{T}{t}-\gamma-1,
\label{eq:Cxxm2}\\
C_{zz}(-1)&=L_{zz}\left(1+\frac{2t}{T}\right)+\frac{2t}{T},\quad L_{zz}\equiv\ln\frac{T}{t}-\gamma-\frac12,
\label{eq:Czzm1}\\
C_{zz}(-\tfrac32)&=\frac{16}{3\pi\sqrt b}-2\sqrt\pi+\frac{96}{5\pi}\sqrt b+O(b),
\label{eq:Czzm32}\\
C_{zz}(-2)&=\frac{T}{t}\left[1-\frac{t}{T}\left(\ln\frac{T}{t}-\gamma-\frac32\right)\right].
\label{eq:Czzm2}
\end{align}
For the two power-law singular cases, $\sigma_{zz}$ with $p=-3/2$ and $p=-2$, the energy integral is dominated by $E\sim t\ll T$, where $-\partial f/\partial E$ is constant and equal to $z/[T(1+z)^{2}]$; since $z/(1+z)^{2}=x(1-\tfrac32x+\tfrac76x^{2}+\cdots)$,
\begin{equation}
c_{1}=-\frac32,\qquad c_{2}=\frac76\qquad(\sigma_{zz},\ p=-\tfrac32,-2).
\label{eq:csingpow}
\end{equation}
For the two logarithmic cases, $\sigma_{xx}$ with $p=-2$ and $\sigma_{zz}$ with $p=-1$, each term $(-1)^{k+1}kz^{k}e^{-kE/T}$ from the Fermi derivative contributes $L-\ln k$, so that
\begin{equation}
c_{1}=-\frac32+\frac{2\ln2}{L},\qquad c_{2}=\frac76+\frac{2\ln2-3\ln3}{L},
\label{eq:csinglog}
\end{equation}
with $L=L_{xx}$ or $L_{zz}$, respectively.

\subsubsection{Scaling summary}

The resulting temperature and density dependences are summarized in Table~\ref{tab:asym}. In the nondegenerate regime both conductivities are linear in $n$ at fixed $T$. The in-plane conductivity scales as $T^{p}$ (times $\ln T$ for $p=-2$). The interlayer conductivity scales as $T^{p-1}$ for $p\ge0$, but for all $p\le-1$ it scales as $T^{-2}$ (times $\ln T$ for $p=-1$), independent of $p$: the energy integral is then dominated by $E\sim t$, and the only temperature dependence is that of the Boltzmann weight, $z/T\simeq\EF/T^{2}$. Correspondingly, $\sigma_{zz}\propto t^{3/2}$ for $p=-3/2$ and $\sigma_{zz}\propto t$ for $p=-2$, rather than the naive $t^{2}$.

\begin{table*}[t]
\caption{Asymptotic conductivities for $t\ll\EF$. Degenerate coefficients are at order $(t/\EF)^0$, $\sigma_{ii}/\sigma_{ii}^0=1+b_{ii}(\pi^2/6)(T/\EF)^2+\cdots$, with $b_{xx}=p(p+1)$ and $b_{zz}=p(p-1)$. The finite-$t$ coefficients are given in Eqs.~\eqref{eq:sommxx} and \eqref{eq:sommzz}; band-edge contributions are discussed in Sec.~\ref{sec:edge}. The density dependence at fixed $E_{0}$ is $\sigma_{xx}\propto n^{1+p}$, $\sigma_{zz}\propto n^{p}$. Nondegenerate regime ($T\gg\EF$): Eqs.~\eqref{eq:highTxx} and \eqref{eq:highTzz}, linear in $n$, with $C$ at leading order in $t/T$ (corrections in Eqs.~\eqref{eq:tcorrxx}--\eqref{eq:Czzm2}) and first quantum correction $c_{1}$; the listed $T$ dependence is at fixed $n$. Exact $t\to0$ forms valid for all $T$ are in Eqs.~\eqref{eq:xxp0}--\eqref{eq:zzp32}.}
\label{tab:asym}
\begin{ruledtabular}
\begin{tabular}{ccccccccc}
 & \multicolumn{4}{c}{In plane, $\sigma_{xx}$} & \multicolumn{4}{c}{Interlayer, $\sigma_{zz}$}\\
\cline{2-5}\cline{6-9}
$p$ & $b_{xx}=p(p+1)$ & $C_{xx}$ & $c_{1}^{xx}$ & $T\gg\EF$ & $b_{zz}=p(p-1)$ & $C_{zz}$ & $c_{1}^{zz}$ & $T\gg\EF$\\
\hline
$2$ & $6$ & $6$ & $3/8$ & $nT^{2}$ & $2$ & $2$ & $1/4$ & $nT$\\
$3/2$ & $15/4$ & $15\sqrt\pi/8$ & $\frac12-2^{-5/2}$ & $nT^{3/2}$ & $3/4$ & $3\sqrt\pi/4$ & $\frac12-2^{-3/2}$ & $nT^{1/2}$\\
$1$ & $2$ & $2$ & $1/4$ & $nT$ & $0$ & $1$ & $0$ & $n$ ($T$ indep.)\\
$0$ & $0$ & $1$ & $0$ & $n$ ($T$ indep.) & $0$ & $1$ & $-1/2$ & $n/T$\\
$-1$ & $0$ & $1$ & $-1/2$ & $n/T$ & $2$ & $\ln(T/t)-\gamma-\frac12$ & $-\frac32+\frac{2\ln2}{L_{zz}}$ & $nT^{-2}\ln(T/t)$\\
$-3/2$ & $3/4$ & $\sqrt\pi-\frac{16}{\pi}\sqrt{t/T}$ & $\frac12-\sqrt2$ & $nT^{-3/2}$ & $15/4$ & $\frac{16}{3\pi}\sqrt{T/t}$ & $-3/2$ & $nT^{-2}\,t^{3/2}$\\
$-2$ & $2$ & $\ln(T/t)-\gamma-1$ & $-\frac32+\frac{2\ln2}{L_{xx}}$ & $nT^{-2}\ln(T/t)$ & $6$ & $T/t$ & $-3/2$ & $nT^{-2}\,t$\\
\end{tabular}
\end{ruledtabular}
\end{table*}

\subsection{Band-edge contributions}
\label{sec:edge}

For the singular cases, the contribution of the miniband bottom, $E\lesssim4t$, is exponentially small in the degenerate regime but multiplied by a large factor, $(\EF/t)^{-p-1}$ or $\ln(T/t)$, which is missed by the Sommerfeld expansion. When $t\ll T$, the occupation factor is essentially constant over the band-edge region, $-\partial f/\partial E\simeq(-\partial f/\partial E)_{E=0}=[4T\cosh^{2}(\mu/2T)]^{-1}$, and the band-edge contribution is
\begin{align}
\frac{\sigma^{\rm edge}}{\sigma^{0}}&=\frac{\EF\,\mathcal{K}_{p}}{4T\cosh^{2}(\mu/2T)}\nonumber\\
&\simeq\frac{\EF}{T}\,e^{-\EF/T}\left(1-e^{-\EF/T}\right)\mathcal{K}_{p},
\label{eq:edge}
\end{align}
where the last form holds for $t\to0$ in the exponent [$z/(1+z)^{2}=e^{-\EF/T}(1-e^{-\EF/T})$ with $z=e^{\EF/T}-1$], and where $\mathcal K_{p}$ is determined by the integrals $\int_{0}^{\infty}\dd E\,E^{p}\Phi(E)$ (Appendix~\ref{app:integrals}), convergent for the power-law cases and cutoff-dependent for the logarithmic cases:
\begin{align}
\sigma_{xx},\ p=-2:&\quad \mathcal K=\ln\frac{T}{t}-\gamma-1,
\label{eq:Kxxm2}\\
\sigma_{zz},\ p=-1:&\quad \mathcal K=\ln\frac{T}{t}-\gamma-\frac12,
\label{eq:Kzzm1}\\
\sigma_{zz},\ p=-\tfrac32:&\quad \mathcal K=\frac{16}{3\pi}\left(\frac{\EF}{t}\right)^{1/2},
\label{eq:Kzzm32}\\
\sigma_{zz},\ p=-2:&\quad \mathcal K=\frac{\EF}{t}.
\label{eq:Kzzm2}
\end{align}
For the two logarithmic cases, the additive constants are exact
in the nondegenerate regime. In the degenerate regime, only the
leading logarithm is meaningful, since the logarithmic tail merges
with the Sommerfeld contribution.
For $T\gg\EF$, the leading band-edge approximation,
Eq.~\eqref{eq:edge}, becomes
$(\EF/T)^2(1-\tfrac32\EF/T)\mathcal K_p$,
recovering the leading terms of
Eqs.~\eqref{eq:Cxxm2}--\eqref{eq:Czzm2}, including
$c_1=-3/2$. The logarithmic cases have the additional
$2\ln2/L$ correction in Eq.~\eqref{eq:csinglog}, arising
from the energy dependence of the logarithm.
For $T\ll\EF$, the edge term adds approximately to the
Sommerfeld result,
$\sigma\simeq\sigma^{\rm Somm}+\sigma^{\rm edge}$.
For $\sigma_{zz}$ with $p=-2$, $T=0.2\,\EF$, and
$t=0.01\,\EF$, the Sommerfeld series through $T^4$ gives
$1.65\,\sigma_{zz}^0$, while Eq.~\eqref{eq:edge} with
$\mu=\mu(T)$ gives $3.03\,\sigma_{zz}^0$.
Their sum, $4.68\,\sigma_{zz}^0$, is within $3\%$ of the
numerical result $4.81\,\sigma_{zz}^0$, despite the edge
term being dominant. With the Sommerfeld series through
$T^6$, the combined result agrees with the numerics to
about $1\%$ at $T=0.1\,\EF$ and better than $10^{-4}$
at $T\le0.05\,\EF$.

Physically, the edge term arises from thermally excited holes
near the miniband bottom, where electrons have a 3D anisotropic
parabolic dispersion and, for $p<0$, long lifetimes.
For the power-law singularities, it overtakes the $T^2$
Sommerfeld correction above a crossover temperature $T^*$
estimated by
$(\EF/T^*)e^{-\EF/T^*}\mathcal K_p
\simeq(\pi^2/6)b_{ii}(T^*/\EF)^2$,
with $b_{xx}=p(p+1)$ and $b_{zz}=p(p-1)$
(Table~\ref{tab:asym}).
This gives $T^*\approx0.1\,\EF$ for $\sigma_{zz}$ with
$p=-2$ and $t=0.01\,\EF$, and parametrically
$T^*\sim\EF/\ln(\EF/t)$.
Exponential activation followed by high-temperature decay
produces a large maximum in $\sigma_{zz}(T)$ at
$T\sim\EF$ for $p\le-1$, and weaker maxima in
$\sigma_{xx}(T)$ for $p=-2$ and $-3/2$
(Sec.~\ref{sec:numerics}).
The normalized interlayer peak height grows as
$(\EF/t)^{-p-1}$ for $p=-3/2,-2$ and logarithmically
for $p=-1$, making it a sensitive probe of the band-bottom scattering.

\section{Numerical results}
\label{sec:numerics}

We evaluate the conductivities using both the two-dimensional
$(\theta,\varepsilon)$ integrals,
Eqs.~\eqref{eq:sxx2d}--\eqref{eq:szz2d}, and the
one-dimensional transport-function expressions,
Eqs.~\eqref{eq:sxx1d}--\eqref{eq:szz1d}, with the chemical
potential determined self-consistently from Eq.~\eqref{eq:density}.
The methods agree within $10^{-9}$ at the tested temperatures;
comparisons with analytical limits appear in
Appendix~\ref{app:validation}. For the singular exponents,
departures from the local Sommerfeld approximation reflect
the growing band-edge contribution omitted by an expansion
around the Fermi surface (Sec.~\ref{sec:edge}).
We use $t=0.01\,\EF$ at fixed density and
$t=0.01\,\EF(n_0)$ for the density sweeps.

\begin{figure*}[t]
\includegraphics[width=\textwidth]{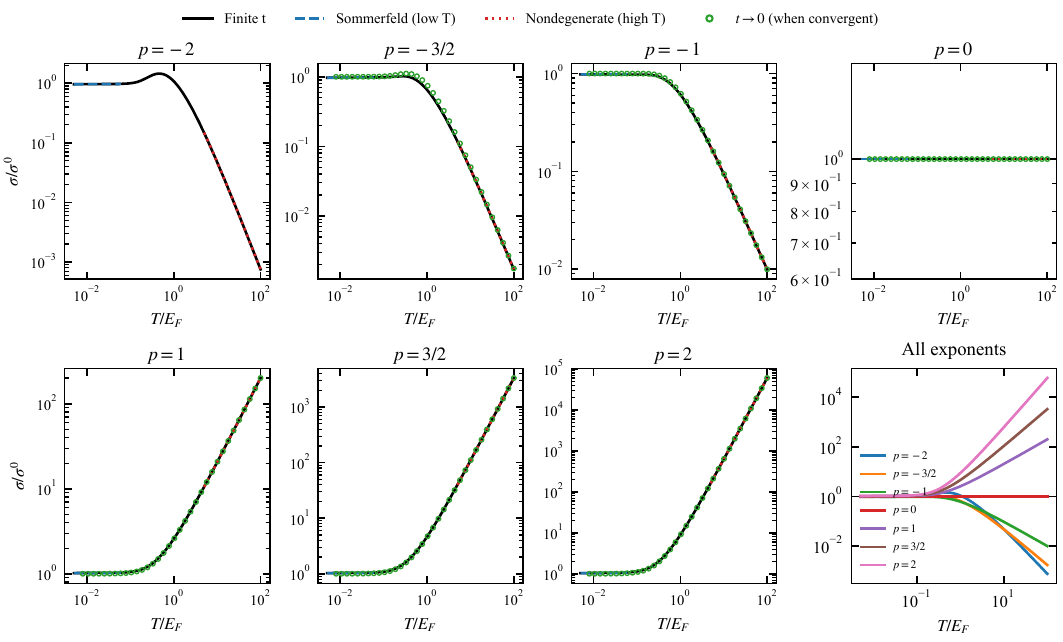}
\caption{In-plane conductivity $\sigma_{xx}/\sigma_{xx}^{0}$ versus $T/\EF$ at fixed density, $t=0.01\,\EF$, for $\tau(E)=\tau(\EF)(E/\EF)^{p}$. Solid: numerical. Dashed: degenerate asymptote, Eq.~\eqref{eq:sommxx} (to $T^{2}$), shown at low temperature. Dotted: nondegenerate asymptote, Eq.~\eqref{eq:highTxx}, shown at high temperature. Circles: exact $t\to0$ result, Eq.~\eqref{eq:exactxx} (not defined for $p=-2$). The last panel collects all seven models. For $p=0$ the conductivity is exactly $T$ independent. For $p=-3/2$, the shallow maximum is $1.02\,\sigma_{xx}^{0}$ at $T=0.27\,\EF$.}
\label{fig:sxxT}
\end{figure*}

\begin{figure*}[t]
\includegraphics[width=\textwidth]{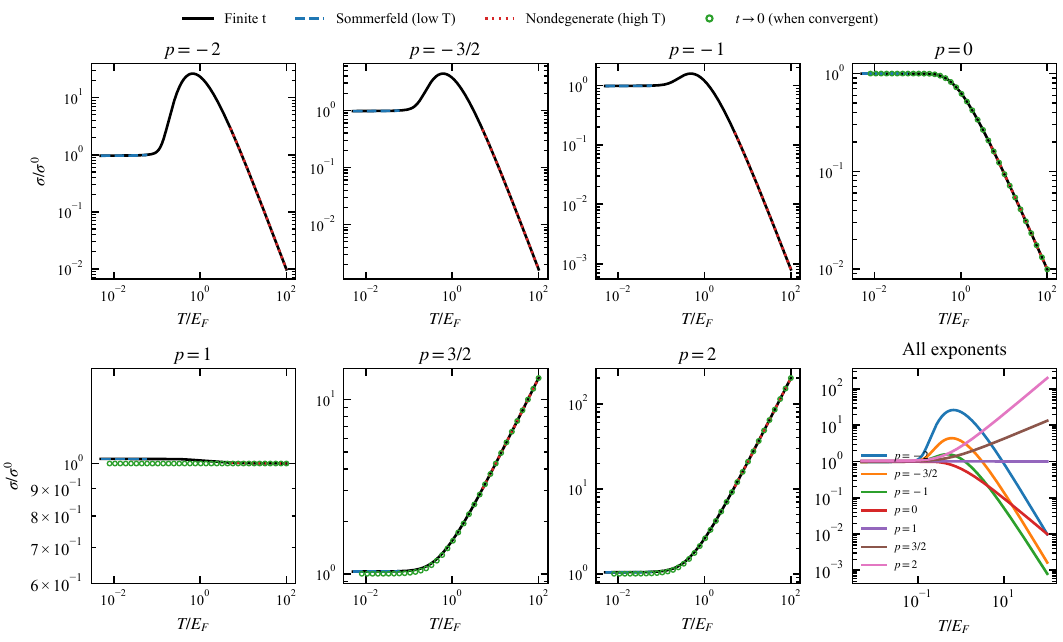}
\caption{Interlayer conductivity $\sigma_{zz}/\sigma_{zz}^{0}$ versus $T/\EF$; same conventions as Fig.~\ref{fig:sxxT}, with Eqs.~\eqref{eq:sommzz}, \eqref{eq:highTzz}, and \eqref{eq:exactzz} (the latter defined only for $p>-1$). For $p=1$ the conductivity is approximately $T$ independent. For $p\le-1$ the large maximum near $T\sim\EF$ is due to carriers at the miniband bottom, Eq.~\eqref{eq:edge}. For $p=-3/2$ and $-1$, the maxima occur at $T=0.61\,\EF$ and $0.48\,\EF$, with heights $4.40\,\sigma_{zz}^{0}$ and $1.58\,\sigma_{zz}^{0}$, respectively.}
\label{fig:szzT}
\end{figure*}

Figures~\ref{fig:sxxT} and \ref{fig:szzT} show
$\sigma_{xx}(T)/\sigma_{xx}^{0}$ and
$\sigma_{zz}(T)/\sigma_{zz}^{0}$ at fixed density for all
seven exponents over more than four decades of $T/\EF$.
The curves are compared with the degenerate
[Eqs.~\eqref{eq:sommxx}, \eqref{eq:sommzz}] and nondegenerate
[Eqs.~\eqref{eq:highTxx}, \eqref{eq:highTzz}] asymptotes and,
where defined, the exact $t\to0$ forms,
Eqs.~\eqref{eq:exactxx} and \eqref{eq:exactzz}.
Selected numerical values appear in Appendix~\ref{app:validation}.

The interlayer conductivity exhibits a much more pronounced maximum than the in-plane conductivity. For $p=-2$ and $t=0.01\,\EF$, the interlayer conductivity reaches $26.3\,\sigma_{zz}^{0}$ at $T=0.67\,\EF$, whereas the in-plane maximum is $1.45\,\sigma_{xx}^{0}$ at $T=0.46\,\EF$. For $p\ge0$ the conductivities are monotonic in $T$.

\begin{figure*}[t]
\includegraphics[width=\textwidth]{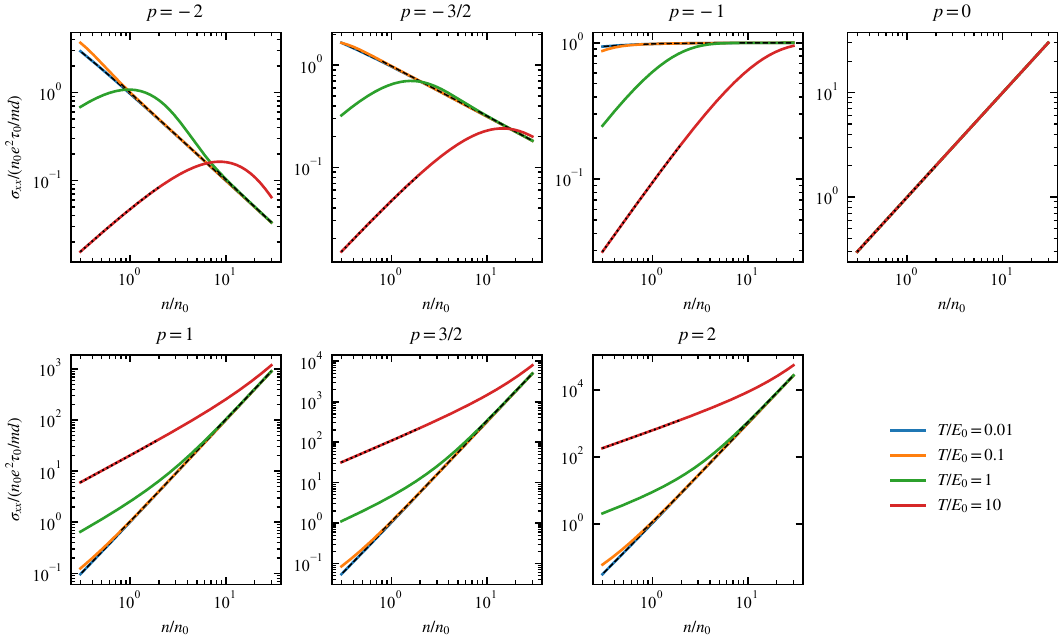}
\caption{In-plane conductivity versus 2D density per layer at fixed $T/E_{0}=0.01$, $0.1$, $1$, $10$, with $E_{0}=\EF(n_{0})$, $\tau(E)=\tau_{0}(E/E_{0})^{p}$, and $t=0.01\,E_{0}$. Dashed: finite-$t$ degenerate asymptote ($\propto n^{1+p}$ with $T^{2}$ correction). Dotted: nondegenerate asymptote ($\propto n$).}
\label{fig:sxxn}
\end{figure*}

\begin{figure*}[t]
\includegraphics[width=\textwidth]{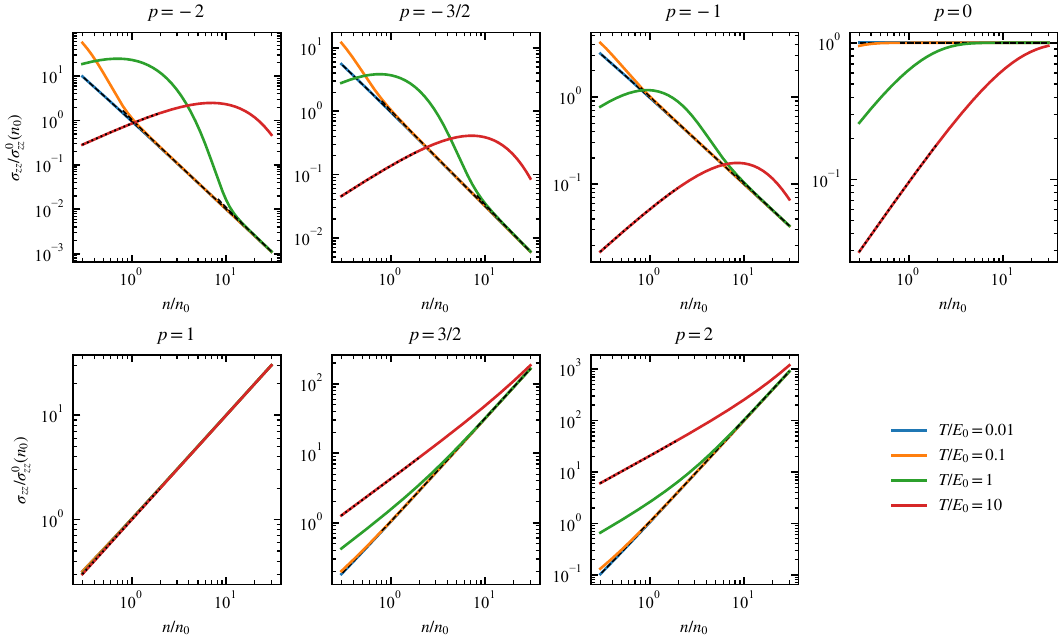}
\caption{Interlayer conductivity versus 2D density per layer; same conventions as Fig.~\ref{fig:sxxn}. The degenerate asymptote is $\propto n^{p}$.}
\label{fig:szzn}
\end{figure*}

The density dependence at fixed temperature is shown in Figs.~\ref{fig:sxxn} and \ref{fig:szzn} for $T/E_{0}=0.01$, $0.1$, $1$, and $10$ over two decades of density, $0.3\le n/n_{0}\le30$. The limiting density dependences are the nondegenerate behavior $\sigma\propto n$ at low density ($\EF\ll T$) and the degenerate behavior $\sigma_{xx}\propto n^{1+p}$, $\sigma_{zz}\propto n^{p}$ at high density ($\EF\gg T$). For $p<0$ the degenerate $\sigma_{zz}$ \emph{decreases} with increasing density, and for $p<-1$ so does $\sigma_{xx}$; this leads to maxima in $\sigma(n)$ at $\EF\sim T$. For $p=1$ the interlayer conductivity is $\sigma_{zz}\propto n$ at all $T$ to leading order in $t/\EF$, and for $p=0$ the in-plane conductivity is $\sigma_{xx}\propto n$ at all $T$, as follows from Eqs.~\eqref{eq:xxp0} and \eqref{eq:zzp1}.

\section{Thermal conductivity and the Wiedemann--Franz law}
\label{sec:WF}

\subsection{Transport coefficients and the Lorenz ratio}

The Wiedemann--Franz (WF) law states that the ratio of the electronic thermal conductivity $\kappa$ to the electrical conductivity $\sigma$ of a metal is $\kappa/(\sigma T)=L_{0}$, with the universal Sommerfeld value (in SI units)
\begin{equation}
L_{0}=\frac{\pi^{2}}{3}\left(\frac{\kB}{e}\right)^{2}=2.443\times10^{-8}\ \mathrm{W\,\Omega\,K^{-2}}.
\label{eq:L0}
\end{equation}
It is a consequence of elastic scattering and a sharp Fermi surface~\cite{AshcroftMermin,Ziman1960,ChesterThellung1961}, and its failure is a sensitive diagnostic of inelastic scattering, hydrodynamics, or nontrivial energy structure of the transport~\cite{PrincipiVignale2015,Crossno2016,MahanSofo1996}. In particular, the WF law is not a unique hallmark of Fermi-liquid behavior: it is violated at finite temperature in an ordinary Fermi liquid of electrons scattered by acoustic phonons and impurities in both 2D and 3D~\cite{LavasaniBulmashDasSarma2019}, and gapped bipolar diffusive transport has been proposed as a possible explanation~\cite{TuDasSarma2023} of the Lorenz-ratio peak observed in graphene~\cite{Crossno2016}. Here we determine the extent of validity of the WF law in both directions of the superlattice for all seven scattering models. We retain $\kB=1$ in the formulas; $S$, $\kappa$, and $\mathcal L$ acquire factors $\kB$, $\kB$, and $\kB^2$, respectively, in SI units. We continue to use a single energy-dependent relaxation time for charge and heat transport, as appropriate for elastic scattering; inelastic effects, which produce additional violations even in a Fermi liquid~\cite{LavasaniBulmashDasSarma2019}, are outside the present scope.

With a temperature gradient in addition to the electric field, the linearized Boltzmann equation gives the charge and heat currents $j_{i}=L^{(11)}_{ii}\mathcal E_{i}+L^{(12)}_{ii}(-\partial_{i}T)$, $j^{Q}_{i}=L^{(21)}_{ii}\mathcal E_{i}+L^{(22)}_{ii}(-\partial_{i}T)$, expressed through the transport moments~\cite{AshcroftMermin,Ziman1960}
\begin{equation}
K_{n}^{ii}=P_{ii}\int_{0}^{\infty}\dd E\,\tau(E)\,\Phi_{ii}(E)\,(E-\mu)^{n}\left(-\frac{\partial f}{\partial E}\right),
\label{eq:Kn}
\end{equation}
with $P_{xx}=N_{0}/md$ and $P_{zz}=2N_{0}t^{2}d/\hbar^{2}$ [cf.~Eqs.~\eqref{eq:sxx1d} and \eqref{eq:szz1d}]. The electrical conductivity, thermopower, and open-circuit thermal conductivity are
\begin{align}
\sigma_{ii}&=e^{2}K^{ii}_{0},\qquad S_{ii}=-\frac{1}{eT}\frac{K^{ii}_{1}}{K^{ii}_{0}},\nonumber\\
\kappa_{ii}&=\frac{1}{T}\left[K^{ii}_{2}-\frac{(K^{ii}_{1})^{2}}{K^{ii}_{0}}\right],
\label{eq:kappa}
\end{align}
where the subtracted term is the Peltier (convective) contribution that is cancelled by the thermoelectric field under open-circuit conditions~\cite{Goldsmid2010}. The Lorenz ratio $\mathcal L_{ii}\equiv\kappa_{ii}/(\sigma_{ii}T)$ is therefore
\begin{equation}
\begin{aligned}
\frac{\mathcal L_{ii}}{L_0}&=\frac{3}{\pi^2T^2}
\left[\frac{K_2^{ii}}{K_0^{ii}}-\left(\frac{K_1^{ii}}{K_0^{ii}}\right)^2\right]\\
&=\frac{3}{\pi^2}\frac{\operatorname{Var}_{W_{ii}}(E)}{T^2} ,
\end{aligned}
\label{eq:Lorenz}
\end{equation}
where $\mathrm{Var}_{W}(E)$ is the variance of the electron energy with respect to the normalized weight
\begin{equation}
W_{ii}(E)=\frac{\tau(E)\Phi_{ii}(E)(-\partial f/\partial E)}{\int\dd E'\,\tau(E')\Phi_{ii}(E')(-\partial f/\partial E')},
\label{eq:weight}
\end{equation}
i.e., the energy distribution of the \emph{current-carrying} electrons. Equation~\eqref{eq:Lorenz} is exact within the Boltzmann theory and provides the physical key to all results below: the WF law holds if and only if the energy variance of the current-carrying electrons equals $(\pi^{2}/3)T^{2}$, which is the variance of the untruncated Fermi window. The lower band edge can also change this variance and hence the Lorenz ratio. Note that $\mathcal L_{ii}$ is independent of the overall scale of $\tau$ and of the prefactor $P_{ii}$, so that it depends only on $T/\EF$, $t/\EF$, and $p$; the density and mass enter only through $\EF=\pi\hbar^{2}n/m$.

\subsection{Degenerate regime: validity of the WF law and its leading corrections}
\label{sec:WFdegenerate}

For $T\ll\EF$, the Sommerfeld expansion uses the moments of the untruncated Fermi window, with $x=E-\mu$:
\begin{equation*}
\begin{aligned}
\int_{-\infty}^{\infty} x^{2k}\left(-\frac{\partial f}{\partial x}\right)\dd x
&=2\left(1-2^{1-2k}\right)\\
&\quad\times\zeta(2k)(2k)!\,T^{2k}.
\end{aligned}
\end{equation*}
For $k=1,2,3$, these moments are $\pi^{2}T^{2}/3$, $7\pi^{4}T^{4}/15$, and $31\pi^{6}T^{6}/21$. With $G_{ii}\equiv\tau\Phi_{ii}$ and its derivatives evaluated at $\mu_{0}$, the expansion gives
\begin{align}
K_{0}&\propto G+\frac{\pi^{2}}{6}T^{2}G''+\frac{7\pi^{4}}{360}T^{4}G''''+\cdots,\nonumber\\
K_{1}&\propto\frac{\pi^{2}}{3}T^{2}G'+\frac{7\pi^{4}}{90}T^{4}G'''+\cdots,
\label{eq:Kdeg}\\
K_{2}&\propto\frac{\pi^{2}}{3}T^{2}G+\frac{7\pi^{4}}{30}T^{4}G''+\cdots,\nonumber
\end{align}
and hence
\begin{equation}
\frac{\mathcal L_{ii}}{L_{0}}=1+\pi^{2}T^{2}\left[\frac{8}{15}\frac{G''_{ii}}{G_{ii}}-\frac{1}{3}\left(\frac{G'_{ii}}{G_{ii}}\right)^{2}\right]+O(T^{4}).
\label{eq:WFdeg}
\end{equation}
Thus the WF law is \emph{exact} as $T\to0$ in both directions, for every $p$, and for every $t<\EF/2$: at $T=0$ the current-carrying window collapses onto the Fermi surface, where $\tau\Phi$ is smooth and elastic scattering cannot distinguish heat from charge. The first correction has two physically distinct parts: the curvature term $G''$, which describes the reweighting of the Fermi window by the energy dependence of $\tau\Phi$, and the negative slope term $-(G')^{2}$, which is the Peltier subtraction in Eq.~\eqref{eq:kappa}.

For the power law with $t\to0$, $G_{xx}\propto E^{p+1}$ and $G_{zz}\propto E^{p}$. Writing $G\propto E^{s}$ with
\begin{equation}
s_{xx}=p+1,\qquad s_{zz}=p,
\label{eq:sdef}
\end{equation}
we obtain the local Sommerfeld expansion, including the next order,
\begin{equation}
\frac{\mathcal L_{ii}}{L_{0}}=1+\lambda_{2}(s)\left(\frac{T}{\EF}\right)^{2}+\lambda_{4}(s)\left(\frac{T}{\EF}\right)^{4}+O(T^{6}),
\label{eq:WFpower}
\end{equation}
\begin{align}
\lambda_{2}(s)&=\frac{\pi^{2}}{15}\,s\,(3s-8),
\label{eq:lambda2}\\
\lambda_{4}(s)&=\frac{2\pi^{4}}{315}\,s\,(s-1)(s-3)(5s-52).
\label{eq:lambda4}
\end{align}
For $\sigma_{zz}$ these results are exact in $t$ upon replacing $\EF\to\mu_{0}=\EF+2t$, since $\Phi_{zz}=1$ above the miniband. For $\sigma_{xx}$, with $G_{xx}=\tau(E)(E-2t)$, Eq.~\eqref{eq:WFdeg} gives exactly~in~$t$
\begin{align}
\frac{\mathcal L_{xx}}{L_{0}}-1&=\pi^{2}T^{2}\bigg[\frac{8}{15}\left(\frac{p(p-1)}{\mu_{0}^{2}}+\frac{2p}{\mu_{0}\EF}\right)\nonumber\\
&\quad-\frac13\left(\frac{p}{\mu_{0}}+\frac{1}{\EF}\right)^{2}\bigg]+O(T^4).
\label{eq:WFxxt}
\end{align}
The coefficients for the fourteen cases are listed in Table~\ref{tab:WF}. Several features stand out:

(i) For $s=0$ ($\sigma_{xx}$ with $p=-1$ and $\sigma_{zz}$ with $p=0$) all power-law corrections vanish, and the WF law holds up to exponentially small terms, $\mathcal L/L_{0}=1+O(e^{-\EF/T})$ (up to $O(tT^{2}/\EF^{3})$ for $\sigma_{xx}$), because the transport weight is exactly $-\partial f/\partial E$.

(ii) For $s=1$ ($\sigma_{xx}$ with $p=0$ and $\sigma_{zz}$ with $p=1$) the series terminates:
\begin{equation}
\frac{\mathcal L_{xx}(p=0)}{L_{0}}=1-\frac{\pi^{2}}{3}\frac{T^{2}}{\EF^{2}},\qquad\frac{\mathcal L_{zz}(p=1)}{L_{0}}=1-\frac{\pi^{2}}{3}\frac{T^{2}}{\mu_{0}^{2}},
\label{eq:WFs1}
\end{equation}
up to exponentially small terms. The first of these is exact in $t$, because by the sum rule~\eqref{eq:sumrule} $K_{0}^{xx}\propto\EF$ exactly while $G_{xx}=E-2t$ is linear. Remarkably, therefore, the in-plane WF law fails at order $T^{2}$ precisely in the model ($p=0$) in which $\sigma_{xx}$ itself is exactly $T$ independent: the heat current is sensitive to the linear energy dependence of $\Phi_{xx}$, which produces a thermopower and hence a Peltier subtraction, while $\sigma_{xx}$ is not.

(iii) The competition between the curvature term and Peltier subtraction gives $\lambda_{2}<0$, i.e., $\mathcal L<L_{0}$, for $0<s<8/3$, and $\lambda_{2}>0$, i.e., $\mathcal L>L_{0}$, for $s<0$ or $s>8/3$. Peltier subtraction reduces the Lorenz ratio, while positive curvature can enhance it by broadening the energy distribution of the carriers. The largest coefficient, $\lambda_{2}=28\pi^{2}/15\approx18.4$, occurs for $\sigma_{zz}$ with $p=-2$.

(iv) Since $\mathcal L_{xx}(p)$ and $\mathcal L_{zz}(p+1)$ have the same $s$, the identity~\eqref{eq:identity} extends to the Lorenz ratio: $\mathcal L_{xx}(p)=\mathcal L_{zz}(p+1)$ for $t\to0$. The finite-temperature Lorenz ratio is thus generally \emph{anisotropic}: $\mathcal L_{xx}\neq\mathcal L_{zz}$ for the same scattering mechanism, $\mathcal L_{xx}-\mathcal L_{zz}\simeq[\lambda_{2}(p+1)-\lambda_{2}(p)]L_{0}T^{2}/\EF^{2}=(\pi^{2}/15)(6p-5)L_{0}T^{2}/\EF^{2}$.

\subsection{Exact Lorenz ratio for all temperatures}

To leading order in $t/\EF$ and for $s>-1$ the moments are complete Fermi--Dirac integrals [Eq.~\eqref{eq:FDmoment}], and since the variance is independent of the reference energy, Eq.~\eqref{eq:Lorenz} can be evaluated with $E$ in place of $E-\mu$:
\begin{align}
\frac{\mathcal L}{L_{0}}&=\frac{3}{\pi^{2}}\bigg[(s+2)(s+1)\frac{\F_{s+2}(z)}{\F_{s}(z)}\nonumber\\
&\quad-(s+1)^{2}\left(\frac{\F_{s+1}(z)}{\F_{s}(z)}\right)^{2}\bigg],
\label{eq:WFexact}
\end{align}
with $z=e^{\EF/T}-1$. This single formula describes the WF ratio at all temperatures for the ten cases with $s>-1$ (the four singular cases, $s\le-1$, are treated below), and reproduces Eq.~\eqref{eq:WFpower} for $z\to\infty$. For $s=0$ it reduces to $\mathcal L/L_{0}=(3/\pi^{2})[2\F_{2}(z)/\F_{0}(z)-(\EF/T)^{2}/\F_{0}(z)^{2}]$ with $\F_{0}=1-e^{-\EF/T}$, which equals $1$ up to exponentially small corrections in the degenerate regime by the inversion formula for $\F_{2}$.

\subsection{Nondegenerate regime: classical Lorenz number}

For $T\gg\EF$ the weight becomes a gamma distribution, $W\propto E^{s}e^{-E/T}$, whose variance is $(s+1)T^{2}$. Expanding Eq.~\eqref{eq:WFexact} in $z$ we find
\begin{equation}
\frac{\mathcal L}{L_{0}}=\frac{3(s+1)}{\pi^{2}}\left[1+(2-s)\,2^{-(s+2)}\,\frac{\EF}{T}\right],
\label{eq:WFclassical}
\end{equation}
for $s>-1$, i.e., the classical (nondegenerate) Lorenz number in SI units $\mathcal L=(s+1)(\kB/e)^{2}$. The Fermi window is replaced by a Boltzmann tail whose relative energy spread depends on how $\tau\Phi$ weights the carriers. The individual classical limits are listed in Table~\ref{tab:WF}. The limiting value exceeds $L_0$ only for $s>\pi^{2}/3-1\approx2.29$, which among the models studied here occurs for $\sigma_{xx}$ with $p=\tfrac32$ and $2$. For $\sigma_{xx}$ with $p=-3/2$ the leading miniband correction is again anomalously large,
\begin{equation}
\frac{\mathcal L_{xx}(-\tfrac32)}{L_{0}}=\frac{3}{2\pi^{2}}\left[1+\frac{8}{\pi^{3/2}}\left(\frac{t}{T}\right)^{1/2}\right]\left[1+\frac{5}{4\sqrt2}\frac{\EF}{T}\right].
\label{eq:WFxxm32}
\end{equation}

\subsection{Singular cases: energy filtering and collapse of the Lorenz ratio}

For $\sigma_{zz}$ with $p\le-1$ and $\sigma_{xx}$ with $p=-2$ the transport weight in the nondegenerate regime is enhanced near the miniband bottom, while higher energy moments sample a broader range (Sec.~\ref{sec:edge}). Evaluating the moments with the regularized integrals of Appendix~\ref{app:integrals} we find:
\begin{align}
\frac{\mathcal L_{zz}(-2)}{L_{0}}&=\frac{3}{\pi^{2}}\frac{t}{T}\,e^{\EF/T}+O\!\left[(t/T)^2\ln^2(T/t)\right],
\label{eq:WFzzm2}\\
\frac{\mathcal L_{zz}(-\tfrac32)}{L_{0}}& \simeq \frac{9}{32\sqrt\pi}\left(\frac{t}{T}\right)^{1/2}\frac{(1+z)^{2}}{z}\,\F_{1/2}(z),
\label{eq:WFzzm32}\\
\frac{\mathcal L_{zz}(-1)}{L_{0}}& \simeq \frac{3}{\pi^{2}}\left[\frac{1}{D_{zz}}-\frac{1-\EF/T}{D_{zz}^{2}}\right],
\label{eq:WFzzm1}\\
\frac{\mathcal L_{xx}(-2)}{L_{0}}& \simeq \frac{3}{\pi^{2}}\left[\frac{1}{D_{xx}}-\frac{1-\EF/T}{D_{xx}^{2}}\right].
\label{eq:WFxxm2}
\end{align}
These are nondegenerate, small-$t/T$ approximations. Here $z=e^{\EF/T}-1$, $L_{xx}$ and $L_{zz}$ are the logarithms of Eqs.~\eqref{eq:Cxxm2} and \eqref{eq:Czzm1}, and $D=L(1-\tfrac32\EF/T)+2\ln2\,\EF/T$. To leading order,
\begin{align}
\frac{\mathcal L_{zz}(-2)}{L_{0}}&\simeq\frac{3}{\pi^{2}}\frac{t}{T},\qquad
\frac{\mathcal L_{zz}(-\tfrac32)}{L_{0}}\simeq\frac{9}{32\sqrt\pi}\sqrt{\frac{t}{T}},\nonumber\\
\frac{\mathcal L}{L_{0}}&\simeq\frac{3}{\pi^{2}}\frac{L-1}{L^{2}},
\label{eq:WFsingleading}
\end{align}
the last for $\sigma_{zz}(p=-1)$ and $\sigma_{xx}(p=-2)$.
The Lorenz ratio thus \emph{collapses} as $t/T\to0$,
scaling as $t/T$ or $(t/T)^{1/2}$ for the power-law
singular cases and $1/\ln(T/t)$ for the logarithmic ones.
This strong WF violation arises because long low-energy
lifetimes outweigh the small velocity weight
$\Phi\propto E^{3/2}$, while the energy variance grows
more slowly than $T^2$ [Eq.~\eqref{eq:Lorenz}].
The mechanism resembles the ``energy filtering'' identified
by Mahan and Sofo as optimal for thermoelectric conversion
through a narrow transport
distribution~\cite{MahanSofo1996}.
Narrower minibands therefore produce larger violations
(Fig.~\ref{fig:WFt}).

\subsection{Degenerate singular cases: two-window enhancement}

In the degenerate regime, the same band-edge carriers produce
the \emph{opposite} effect. Their contribution to $\sigma$
is exponentially small [Eq.~\eqref{eq:edge}], but carriers
near $E=0$ lie an energy $\mu_0$ below the Fermi level,
giving their contribution to the energy variance a large
weight $\mu_0^2\gg T^2$.
Treating the transport weight as a mixture of the
Fermi-surface window (weight $1$, variance
$\pi^2T^2\mathcal L^{\rm Somm}/3L_0$) and a band-edge
component (relative weight $r$, at $E-\mu\simeq-\mu_0$) gives
\begin{equation}
\frac{\mathcal L}{L_0}\simeq
\frac{\mathcal L^{\rm Somm}/L_0}{1+r}
+\frac{3}{\pi^2}\left(\frac{\mu_0}{T}\right)^2
\frac{r}{(1+r)^2},
\qquad
r=\frac{\sigma^{\rm edge}}{\sigma^{\rm Somm}},
\label{eq:WFbipolar}
\end{equation}
with $\sigma^{\rm edge}$ from Eq.~\eqref{eq:edge} and
$\mathcal L^{\rm Somm}$ from Eq.~\eqref{eq:WFpower}.
The edge component's intrinsic variance and the thermoelectric shift
of the Fermi-surface mean are neglected. For $r\ll1$, the enhancement is
$\delta\mathcal L/L_0\simeq
(3/\pi^2)(\EF/T)^3e^{-\EF/T}\mathcal K_p$.
The energy separation is analogous to bipolar heat
conduction~\cite{Goldsmid2010}, although both components here are
electrons in the same band. This produces a pronounced maximum of
$\mathcal L_{zz}$ at $T\approx0.1$--$0.2\,\EF$.
As $T$ increases, band-edge carriers dominate $\sigma$ ($r\gg1$),
and the Lorenz ratio approaches the collapsed values of
Eq.~\eqref{eq:WFsingleading}. For these singular cases, the WF law
thus fails in two opposite ways.

\subsection{Extent of validity: regimes in $n$, $T$, $t$, and $m$}

\begin{figure*}[t]
\includegraphics[width=\textwidth]{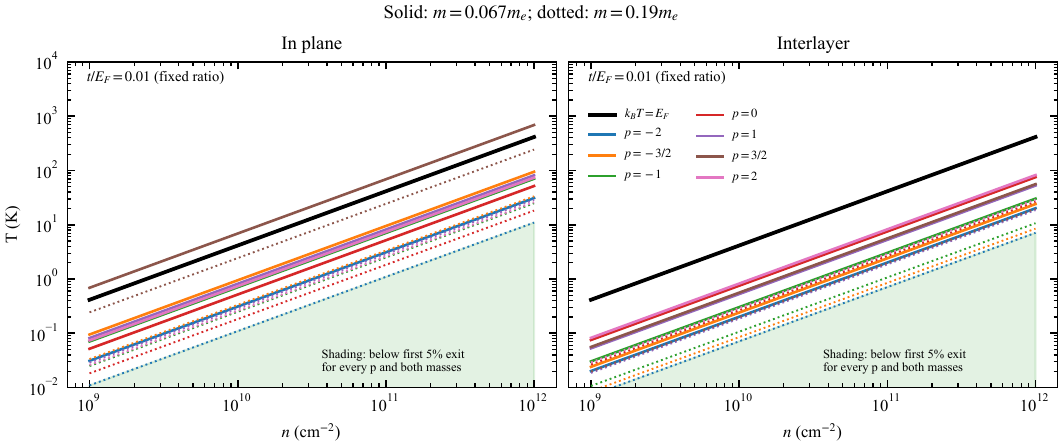}
\caption{Extent of validity of the WF law for a GaAs-like superlattice. Lines: the first temperature at which $|\mathcal L/L_0-1|$ reaches $5\%$, $T_{5\%}=\alpha_p\EF(n)/\kB$ in kelvin, for each scattering model (Table~\ref{tab:WF}), for $m=0.067\,m_{e}$ (solid) and $0.19\,m_{e}$ (dotted). The ratio $t/\EF=0.01$ is fixed as density varies. The thick black line is $\kB T=\EF$ for $m=0.067\,m_e$. In the shaded region the WF law holds to $5\%$ for all seven models and both displayed masses.}
\label{fig:WFmap}
\end{figure*}

Collecting these results, the leading low-temperature estimates for agreement with the WF law within a tolerance $\delta$ are
\begin{equation}
|\lambda_{2}(s)|\left(\frac{T}{\EF}\right)^{2}\lesssim\delta\quad\text{and}\quad\frac{3}{\pi^{2}}\left(\frac{\EF}{T}\right)^{3}e^{-\EF/T}\mathcal K_{p}\lesssim\delta,
\label{eq:WFcriterion}
\end{equation}
the second condition applying only to the singular cases. In physical parameters, the first condition reads
\begin{equation}
T\lesssim\alpha_{p}\,\EF=\alpha_{p}\,\frac{\pi\hbar^{2}n}{m},
\label{eq:WFnm}
\end{equation}
with $\alpha_{p}\simeq(\delta/|\lambda_{2}|)^{1/2}$ when the quadratic correction dominates; the WF law is thus favored by high density and light in-plane mass. The hopping $t$ enters through the finite-$t$ transport functions and the band-edge contributions. The interlayer mass $m_{z}$ has no separate effect at fixed $t/\EF$, because both $\sigma_{zz}$ and $\kappa_{zz}$ scale with the same prefactor $t^{2}d$, which cancels in $\mathcal L_{zz}$. The numerically determined thresholds $\alpha_{p}$ for $\delta=5\%$ are listed in Table~\ref{tab:WF} and denote the first exit from the tolerance on warming from $T=0$. For the nonsingular cases they range from $0.12$ to $0.23$ (except $\sigma_{xx}$ with $p=\tfrac32$, whose small $\lambda_{2}=-\pi^{2}/12$ and classical limit $21/(2\pi^{2})=1.064$ keep it within $5\%$ of $L_{0}$ up to $T\approx1.7\,\EF$), and for the singular cases they are reduced to $0.05$--$0.075$ by the band-edge enhancement (for $\sigma_{zz}$) or the large $\lambda_{2}$. The corresponding validity boundaries for a GaAs-like superlattice ($m=0.067\,m_{e}$, $\EF/\kB=41.5$~K at $n=10^{11}$~cm$^{-2}$) are shown in Fig.~\ref{fig:WFmap}: the WF law holds to $5\%$ for all models below $T\approx2\,\mathrm{K}\times(n/10^{11}\,\mathrm{cm}^{-2})$, and a heavier mass lowers this boundary proportionally.

\subsection{Numerical results}

\begin{figure*}[t]
\includegraphics[width=\textwidth]{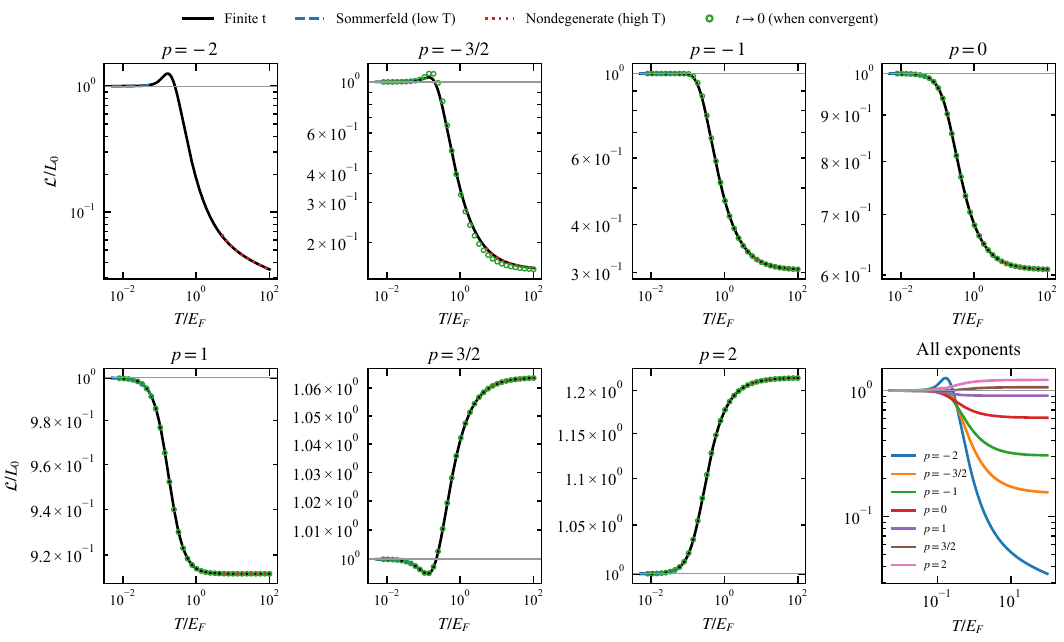}
\caption{In-plane Lorenz ratio $\mathcal L_{xx}/L_{0}$ versus $T/\EF$ for $t=0.01\,\EF$. Solid: numerical. Dashed: finite-$t$ Sommerfeld expansion through $T^4$ at low temperature. Dotted: nondegenerate results, Eqs.~\eqref{eq:WFclassical}, \eqref{eq:WFxxm32}, and \eqref{eq:WFxxm2}. Circles: exact $t\to0$ result, Eq.~\eqref{eq:WFexact}. The horizontal line marks the WF value. For $p=-2$, the maximum is $1.26\,L_{0}$.}
\label{fig:WFxx}
\end{figure*}

\begin{figure*}[t]
\includegraphics[width=\textwidth]{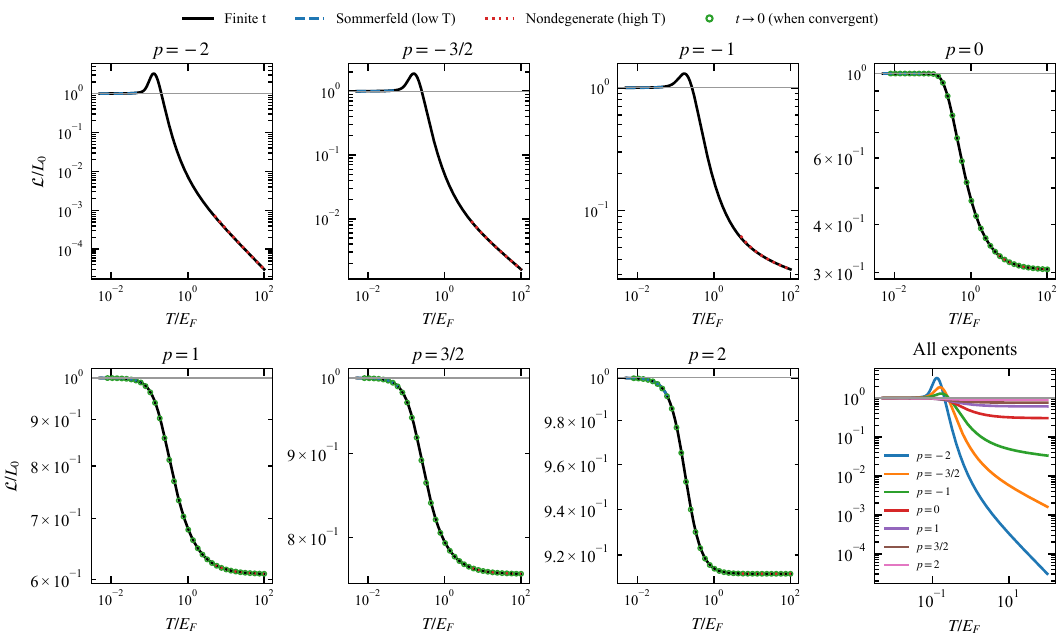}
\caption{Interlayer Lorenz ratio $\mathcal L_{zz}/L_{0}$; same conventions as Fig.~\ref{fig:WFxx}, with Eqs.~\eqref{eq:WFzzm2}--\eqref{eq:WFzzm1} for the singular cases. For $p\le-1$ the Lorenz ratio first rises above $L_{0}$ (two-window enhancement by band-edge carriers) and then collapses (energy filtering). For $p=0$ ($s=0$) the WF law holds up to exponentially small corrections throughout the degenerate regime. At $t=0.01\,\EF$, the maxima for $p=-3/2$ and $-1$ are $1.89\,L_{0}$ and $1.31\,L_{0}$, respectively.}
\label{fig:WFzz}
\end{figure*}

\begin{figure*}[t]
\includegraphics[width=\textwidth]{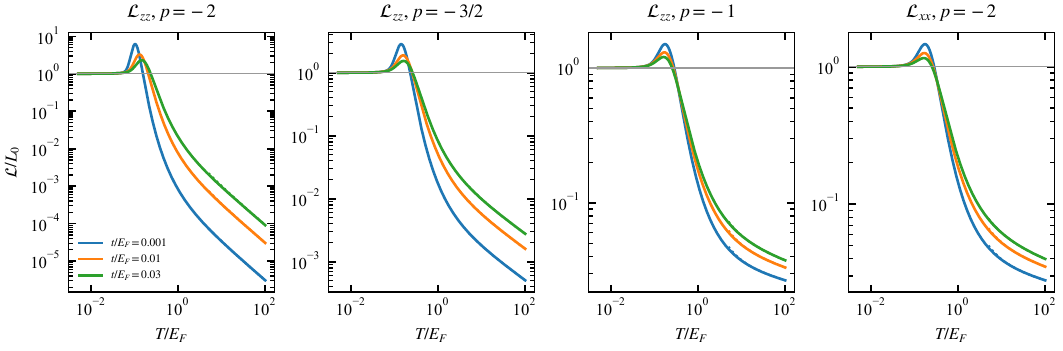}
\caption{Dependence of the WF violation on the miniband width for the singular cases, for $t/\EF=0.001$, $0.01$, $0.03$. Solid: numerical; dotted: nondegenerate results, Eqs.~\eqref{eq:WFzzm2}--\eqref{eq:WFxxm2}. Narrower minibands produce both a larger maximum and a stronger high-temperature collapse [$\propto t/T$, $(t/T)^{1/2}$, $1/\ln(T/t)$].}
\label{fig:WFt}
\end{figure*}

The Lorenz ratios computed numerically from Eqs.~\eqref{eq:Kn}--\eqref{eq:Lorenz} are shown in Figs.~\ref{fig:WFxx} and \ref{fig:WFzz} together with the finite-$t$ Sommerfeld expansion through $T^4$, exact $t\to0$ [Eq.~\eqref{eq:WFexact}], and nondegenerate [Eqs.~\eqref{eq:WFclassical}--\eqref{eq:WFxxm2}] results. At $T=0.1\,\EF$, $\mathcal L_{zz}(p=-2)=2.24\,L_{0}$---more than twice the WF value. The local Sommerfeld expansion underestimates it by $43\%$ because it omits the band-edge contribution. Figure~\ref{fig:WFt} shows the stronger enhancement and high-temperature suppression for narrower minibands. Detailed comparisons with the analytical approximations are given in Appendix~\ref{app:validation}.

\begin{table*}[t]
\caption{Wiedemann--Franz law in the superlattice. $s$: exponent of the transport weight, Eq.~\eqref{eq:sdef}. $\lambda_{2}$, $\lambda_{4}$: degenerate coefficients at leading order in $t/\EF$, Eqs.~\eqref{eq:lambda2} and \eqref{eq:lambda4}. $\mathcal L_{\rm high}/L_0$: leading high-temperature behavior, Eqs.~\eqref{eq:WFclassical} and \eqref{eq:WFsingleading}. $\alpha_{5\%}$: value of $T/\EF$ at which $|\mathcal L/L_{0}-1|$ first reaches $5\%$ (numerical, $t=0.01\,\EF$).}
\label{tab:WF}
\begin{ruledtabular}
\begin{tabular}{c|ccccc|ccccc}
 & \multicolumn{5}{c|}{In plane, $\mathcal L_{xx}$} & \multicolumn{5}{c}{Interlayer, $\mathcal L_{zz}$}\\
$p$ & $s$ & $\lambda_{2}$ & $\lambda_{4}$ & $\mathcal L_{\rm high}/L_0$ & $\alpha_{5\%}$ & $s$ & $\lambda_{2}$ & $\lambda_{4}$ & $\mathcal L_{\rm high}/L_0$ & $\alpha_{5\%}$\\
\hline
$2$ & $3$ & $\pi^{2}/5$ & $0$ & $12/\pi^{2}$ & $0.18$ & $2$ & $-4\pi^{2}/15$ & $8\pi^{4}/15$ & $9/\pi^{2}$ & $0.20$ \\
$3/2$ & $5/2$ & $-\pi^{2}/12$ & $79\pi^{4}/168$ & $\frac{21}{2\pi^{2}}$ & $1.66$ & $3/2$ & $-7\pi^{2}/20$ & $89\pi^{4}/280$ & $\frac{15}{2\pi^{2}}$ & $0.14$ \\
$1$ & $2$ & $-4\pi^{2}/15$ & $8\pi^{4}/15$ & $9/\pi^{2}$ & $0.20$ & $1$ & $-\pi^{2}/3$ & $0$ & $6/\pi^{2}$ & $0.13$ \\
$0$ & $1$ & $-\pi^{2}/3$ & $0$ & $6/\pi^{2}$ & $0.12$ & $0$ & $0$ & $0$ & $3/\pi^{2}$ & $0.18$ \\
$-1$ & $0$ & $0$ & $0$ & $3/\pi^{2}$ & $0.17$ & $-1$ & $11\pi^{2}/15$ & $304\pi^{4}/105$ & $\frac{3}{\pi^{2}}\frac{L-1}{L^{2}}$ & $0.073$ \\
$-3/2$ & $-1/2$ & $19\pi^{2}/60$ & $109\pi^{4}/120$ & $\frac{3}{2\pi^{2}}$ & $0.23$ & $-3/2$ & $5\pi^{2}/4$ & $51\pi^{4}/8$ & $\frac{9}{32\sqrt\pi}\sqrt{t/T}$ & $0.058$ \\
$-2$ & $-1$ & $11\pi^{2}/15$ & $304\pi^{4}/105$ & $\frac{3}{\pi^{2}}\frac{L-1}{L^{2}}$ & $0.075$ & $-2$ & $28\pi^{2}/15$ & $248\pi^{4}/21$ & $\frac{3}{\pi^{2}}\frac{t}{T}$ & $0.049$ \\
\end{tabular}
\end{ruledtabular}
\end{table*}

\section{Discussion}
\label{sec:discussion}

\subsection{Anisotropy}

\begin{figure}[t]
\includegraphics[width=\columnwidth]{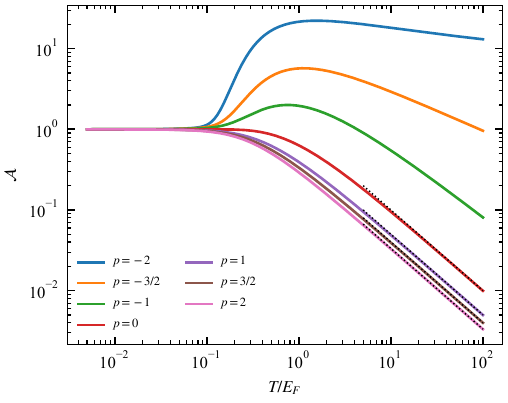}
\caption{Reduced anisotropy $\mathcal A$, Eq.~\eqref{eq:anisotropy}, versus $T/\EF$ for all seven models, $t=0.01\,\EF$. Dotted lines: high-$T$ limit $(\EF/T)/(p+1)$ for $p>-1$, Eq.~\eqref{eq:AhighT}.}
\label{fig:aniso}
\end{figure}

We define the reduced anisotropy
\begin{equation}
\mathcal A\equiv\frac{\sigma_{zz}/\sigma_{xx}}{(m/m_{z})(t/\EF)},
\label{eq:anisotropy}
\end{equation}
which equals unity at $T=0$ [Eq.~\eqref{eq:ratio0}]. From Eqs.~\eqref{eq:sommxx0} and \eqref{eq:sommzz0},
\begin{equation}
\mathcal A=1-\frac{\pi^{2}}{3}\,p\,\frac{T^{2}}{\EF^{2}}+\cdots\qquad(T\ll\EF),
\label{eq:AlowT}
\end{equation}
and from Eqs.~\eqref{eq:highTxx}--\eqref{eq:Czzm2},
\begin{equation}
\mathcal A\simeq\frac{\EF}{T}\frac{C_{zz}(p)}{C_{xx}(p)}=\frac{\EF}{T}\,\frac{1}{p+1}\qquad(T\gg\EF,\ p>-1).
\label{eq:AhighT}
\end{equation}
Thus the physical anisotropy decreases in the nondegenerate regime as
$\sigma_{zz}/\sigma_{xx}\simeq(m/m_z)(t/T)/(p+1)$:
the in-plane kinetic energy grows as $T$, while
$\langle v_z^2\rangle$ saturates at the miniband value.
For the singular exponents, the band bottom controls the high-$T$
anisotropy. For $p=-2$,
$\sigma_{zz}/\sigma_{xx}\simeq(m/m_z)/[\ln(T/t)-\gamma-1]$,
giving the anisotropy of a 3D ellipsoidal band up to a logarithm:
interlayer transport is dominated by $E\sim t$, whereas the in-plane
conductivity retains a logarithmic tail. For $p=-3/2$,
$\sigma_{zz}/\sigma_{xx}\simeq
\frac{16}{3\pi^{3/2}}(m/m_z)(t/T)^{1/2}$.
The reduced anisotropy is shown in Fig.~\ref{fig:aniso}.

\subsection{Resistivity}

The resistivities are $\rho_{xx}=1/\sigma_{xx}$ and $\rho_{zz}=1/\sigma_{zz}$, with reference values $\rho_{ii}^{0}\equiv1/\sigma_{ii}^{0}$ ($i=x,z$). At low temperature, to leading order in $t/\EF$, Eqs.~\eqref{eq:sommxx0} and \eqref{eq:sommzz0} show that $\rho(T)$ decreases as $T^{2}$ (an ``insulating-like'' sign of $\dd\rho/\dd T$) for every case with $b_{ii}>0$ in Table~\ref{tab:asym}, while for $\sigma_{xx}$ with $p=-1$ and $\sigma_{zz}$ with $p=0$ the resistivity \emph{increases} exponentially slowly, $\rho=\rho^{0}/(1-e^{-\EF/T})$, and for $\sigma_{xx}$ with $p=0$ and $\sigma_{zz}$ with $p=1$ it is strictly constant. In the nondegenerate regime $\rho_{xx}\propto T^{-p}$ for $p>-2$, while $\rho_{xx}(p=-2)\propto T^2/\ln(T/t)$, and $\rho_{zz}\propto T^{1-p}$ (for $p\ge0$) or $\propto T^{2}$ (for $p\le-1$, up to logarithms). We stress that these temperature dependences are purely kinematic, arising from thermal averaging with a $T$-independent $\tau(E)$. In real 2D and quasi-2D systems the temperature dependence of screening can produce a linear-in-$T$ resistivity at $T\ll\EF$ which can overwhelm the $T^{2}$ Sommerfeld terms found here~\cite{Stern1980,GoldDolgopolov1986,DasSarmaHwang1999,DasSarmaRMP2011}. The present results should be regarded as the baseline to which such effects are added, and they become directly applicable whenever $\tau$ is controlled by $T$-independent scatterers. For quasielastic phonons in the equipartition regime, the explicit temperature-dependent scattering prefactor must also be included.

\subsection{Validity}

\begingroup
\widowpenalty=10000
\looseness=-1
Our results rest on four assumptions.
(i) Coherent miniband transport: the Boltzmann treatment of
$\sigma_{zz}$ requires $\hbar/\tau_q\ll t$ at the energies
carrying transport, where $\tau_q$ is the quantum rather than
transport lifetime~\cite{DasSarmaStern1985}. In the incoherent
limit, a tunneling formula replaces the Boltzmann
expression~\cite{McKenzieMoses1998}.
(ii) Semiclassical transport: $k_F\ell\gg1$, with $\ell$ the
in-plane mean free path, and weak disorder at the energies
carrying each transport moment. Quantum-interference and
localization corrections are neglected~\cite{LeeRamakrishnan1985}.
(iii) A single miniband: $T$ and $\EF$ must be small compared
with the gap to the next miniband. The nondegenerate
asymptotics require $\EF\ll T$ while $T$ remains below this gap.
(iv) Power-law scattering down to the band bottom: for $p<0$,
other scattering mechanisms can cut off the divergence
$\tau\to\infty$, e.g., through Matthiessen's rule
$\tau^{-1}=\tau_p^{-1}+\tau_c^{-1}$, where $\tau_p$ is the
power-law lifetime and $\tau_c$ describes the additional
scattering channel. The band-edge results of
Sec.~\ref{sec:edge} apply when the cutoff energy is small
compared with $t$; otherwise the transport integrals require
the actual lifetime. The sensitivity of the $\sigma_{zz}$
maximum to this cutoff makes it a useful diagnostic of
low-energy scattering.
Finally, the closed-Fermi-surface regime $\EF<2t$ can be
treated with the same transport functions,
Eqs.~\eqref{eq:Phixx} and \eqref{eq:Phizz}, and will be
discussed elsewhere.

\par\endgroup

\section{Possible implications for the layered cuprates}
\label{sec:cuprates}

The high-$T_c$ cuprates are well-known anisotropic layered
metals, with conducting CuO$_2$ planes stacked along the
$c$ axis, corresponding to our $z$ axis.
Although strong correlations and nonparabolic in-plane
dispersion complicate the comparison, we ask which kinematic
results survive, taking $m_z\sim10\,m$ as a representative
interlayer band mass. The discussion remains speculative.

\emph{Fermi-surface topology and resistivity anisotropy.} In overdoped Tl$_{2}$Ba$_{2}$CuO$_{6+\delta}$, angle-dependent magnetoresistance has revealed a coherent, weakly warped cylindrical Fermi surface~\cite{Hussey2003}, i.e., the open, corrugated Fermi surface of our $t\ll\EF$ regime. For such a Fermi surface our $T=0$ result, Eq.~\eqref{eq:ratio0}, combined with Eq.~\eqref{eq:mz}, gives the resistivity anisotropy
\begin{equation}
\frac{\rho_{c}}{\rho_{ab}}=\frac{\sigma_{xx}}{\sigma_{zz}}=\frac{m_{z}}{m}\,\frac{\EF}{t}=2\pi nd^{2}\left(\frac{m_{z}}{m}\right)^{2}.
\label{eq:cuprateaniso}
\end{equation}
Crucially, the anisotropy is \emph{not} the band-mass ratio $m_{z}/m$: the Fermi-surface average differs from the band-bottom curvature, and the effective interlayer transport mass is $m_{zz}^{*}=m_{z}\EF/t$ (Sec.~\ref{sec:T0}). 
With $m_z=10\,m$ and cuprate-like parameters
(in-plane lattice constant $a\approx3.8$~\AA,
interplane spacing $d\approx6.6$~\AA, and an assumed
$n\approx p_h/a^2$ with hole doping
$p_h\approx0.15$--$0.2$), we find
$2\pi nd^2\approx3$, $t/\EF\approx0.03$, and
$\rho_c/\rho_{ab}\sim3\times10^2$, rather than $10$.
If the full large Fermi surface ($n\approx(1+p_h)/a^{2}$) is used instead, the anisotropy is larger still. An open Fermi surface thus amplifies a moderate mass anisotropy into a large transport anisotropy, of the same order as the large $\rho_{c}/\rho_{ab}$ characteristic of the cuprates. In a clean, single-band weak-coupling BCS description at $T=0$~\cite{Kogan2002}, the London penetration depths are controlled by the same Fermi-surface averages, giving \mbox{$\lambda_{c}^{2}/\lambda_{ab}^{2}=m_{zz}^{*}/m$ for this model}.

\emph{Temperature dependence of the anisotropy.} Within coherent Boltzmann transport with a common scattering time, Eq.~\eqref{eq:AlowT} gives the leading low-temperature anisotropy factor for a smooth power-law lifetime away from the miniband edges, $1-(\pi^{2}/3)p(T/\EF)^{2}$, whose temperature-dependent correction is negligible for $T\ll\EF$. In many cuprates, particularly underdoped ones, $\rho_{c}$ and $\rho_{ab}$ nonetheless have qualitatively different temperature dependences. Our results do not rule out energy-dependent elastic scattering; additional possibilities include incoherent interlayer transport~\cite{McKenzieMoses1998}, a strongly momentum-dependent interlayer hopping $t_{\perp}(\bm k)\propto(\cos k_{x}a-\cos k_{y}a)^{2}$~\cite{Andersen1995}, which weights the $c$-axis current toward the antinodal regions where the pseudogap and the strongest scattering reside~\cite{IoffeMillis1998}, or anisotropic (``hot-spot'') scattering.

\emph{Wiedemann--Franz law.} Our analysis shows that, for elastic scattering, the WF law holds exactly as $T\to0$ in both the $ab$ plane and along $c$, with kinematic corrections of relative order $(T/\EF)^{2}$ and an anisotropy $\mathcal L_{ab}-\mathcal L_{c}\simeq(\pi^{2}/15)(6p-5)L_{0}(T/\EF)^{2}$ when the smooth Fermi-surface contribution dominates. At $100$ K with $\EF/\kB\sim10^3$ K the squared thermal-energy ratio is $10^{-2}$, before accounting for the coefficients and any band-edge contribution. Reported cuprate WF violations~\cite{Hill2001} and observations of its validity~\cite{Proust2002,Michon2018} require material-specific analysis. A comparison of $\mathcal L_{c}$ and $\mathcal L_{ab}$ measured on the same crystal would provide a sharp test: both approach $L_0$ as $T\to0$ in this elastic model, while finite-temperature anisotropy need not imply incoherence. A van Hove singularity near a cuprate Lifshitz transition is a saddle point, rather than the parabolic band minimum treated here; possible Lorenz-ratio signatures require recalculating the DOS, velocities, and scattering. Whether such signatures are observable and how they are modified by the momentum dependence of $t_{\perp}$ and by correlations remain open questions.

\section{Conclusion}
\label{sec:conclusion}

We have solved the Boltzmann transport problem for a tight-binding
superlattice of 2D parabolic layers with power-law energy-dependent
scattering in the regime $t\ll\EF$. The main results are:
(i) Both conductivities reduce exactly to one-dimensional energy
integrals with closed-form transport functions. To leading order in
$t/\EF$, complete Fermi--Dirac integrals give the full temperature
dependence for the convergent cases.
(ii) An exact sum rule makes the in-plane conductivity temperature
independent for $p=0$ at any $t$. The interlayer conductivity is
temperature independent for $p=1$ to leading order in $t/\EF$.
(iii) For $p=-2$ in plane and $p\le-1$ out of plane, long-lived
band-edge carriers can dominate transport. The miniband regularizes
the singular low-energy integrals, yielding logarithmic or anomalous
power-law hopping dependences and a pronounced interlayer conductivity
maximum at $T\sim\EF$.
(iv) The Lorenz ratio measures the energy variance of the
current-carrying electrons. The WF law is recovered in both directions
as $T\to0$ at fixed $t$, with anisotropic finite-temperature corrections.
(v) Band-edge carriers can first enhance the interlayer Lorenz ratio
through contributions from separated energy ranges within the same
band and then suppress it at high temperature through energy filtering.

As discussed in Sec.~\ref{sec:cuprates}, these results may also bear
on the layered cuprates. An open Fermi surface converts a moderate
interlayer mass ratio $m_z/m\sim10$ into a transport anisotropy
$\rho_c/\rho_{ab}=2\pi nd^2(m_z/m)^2\sim10^2$--$10^3$.
Application to a cuprate requires its actual carrier count, band shape,
and scattering; other energy-dependent elastic mechanisms remain
possible. Extending the results to a van Hove saddle point requires
recalculating the transport kernels.

Our results reinforce the importance of detailed, system-dependent
transport theories: even this simple anisotropic layered metal can
exhibit counterintuitive temperature dependence controlled by its band
structure and scattering. A normal metal can show increasing or
nonmonotonic conductivity over some temperature range.
Our analysis treats each scattering exponent $p$ separately, whereas
the dominant resistive scattering process often changes with
temperature. For example, phonon scattering is strongly suppressed
in the low-temperature Bloch--Gr\"uneisen regime, where impurity
scattering and, where kinematically allowed, umklapp electron--electron
scattering can dominate transport~\cite{PalYudsonMaslov2012}.
At higher, system-dependent temperatures, phonons often dominate
the resistivity. The effective energy dependence of the scattering
time can therefore change with temperature, beyond our fixed-$p$
description. Multiple scattering processes may also coexist,
further complicating the resistivity and limiting generic conclusions.
These considerations also bear on the Wiedemann--Franz law, which
is restored as $T\to0$ in a Fermi-liquid metal with finite elastic
impurity scattering, but may fail strongly at finite temperatures
depending on the active scattering
mechanisms~\cite{LavasaniBulmashDasSarma2019}.

\begin{acknowledgments}
This work is supported by the Laboratory for Physical Sciences. The authors acknowledge Anthropic Claude Fable 5 and OpenAI GPT Astra 6 for AI assistance with figure preparation, numerical calculations, and editing. The authors are responsible for the work.
\end{acknowledgments}

\appendix
\raggedbottom

\section{Angular averages and energy integrals}
\label{app:integrals}

With $w=4t\sin^{2}\phi$, $\phi=\theta/2$, and $\sin^{2}\theta=4\sin^{2}\phi\cos^{2}\phi$, the $\theta$ average becomes $\mean{\cdots}=(2/\pi)\int_{0}^{\pi/2}\dd\phi(\cdots)$. Using $\int_{0}^{\pi/2}\ln\sin\phi\,\dd\phi=-(\pi/2)\ln2$, $\int_{0}^{\pi/2}\cos(2k\phi)\ln\sin\phi\,\dd\phi=-\pi/4k$ ($k\ge1$), and $\int_{0}^{\pi/2}\sin^{2a-1}\phi\cos^{2b-1}\phi\,\dd\phi=\tfrac12B(a,b)$, where $B$ is the Euler beta function and $\operatorname{Re}a,\operatorname{Re}b>0$, we find
\begin{align}
\mean{\ln w}&=\ln t, &\mean{\sin^{2}\theta\ln w}&=\tfrac12\ln t+\tfrac14,
\label{eq:A1}\\
\mean{\sqrt w}&=\frac{4\sqrt t}{\pi}, &\mean{\frac{\sin^{2}\theta}{\sqrt w}}&=\frac{4}{3\pi\sqrt t},
\label{eq:A2}\\
\mean{\sin^{2}\theta}&=\tfrac12, &\mean{\frac{\sin^{2}\theta}{w}}&=\frac{1}{2t}.
\label{eq:A3}
\end{align}
The energy integrals needed for $a=w/T\ll1$ are
\begin{align}
\int_{0}^{\infty}\frac{x\,e^{-x}\,\dd x}{(x+a)^{2}}&=(1+a)e^{a}E_{1}(a)-1\nonumber\\
&\simeq-\ln a-\gamma-1,
\label{eq:A4}\\
\int_{0}^{\infty}\frac{e^{-x}\,\dd x}{x+a}&=e^{a}E_{1}(a)\simeq-\ln a-\gamma,
\label{eq:A5}\\
\int_{0}^{\infty}\frac{e^{-x}\,\dd x}{(x+a)^{3/2}}&\simeq\frac{2}{\sqrt a}-2\sqrt\pi,
\label{eq:A6}\\
\int_{0}^{\infty}\frac{e^{-x}\,\dd x}{(x+a)^{2}}&\simeq\frac1a,
\label{eq:A7}\\
\int_{0}^{\infty}\!\Big[\tfrac{x}{(1+x)^{3/2}}-x^{-1/2}\Big]\dd x&=-4,
\label{eq:A8}
\end{align}
where $E_1$ is the exponential integral. Using
Eqs.~\eqref{eq:A1}--\eqref{eq:A8} and
$-\partial f/\partial E\simeq(z/T)e^{-E/T}$ in
Eqs.~\eqref{eq:sxx2d} and \eqref{eq:szz2d} gives
Eqs.~\eqref{eq:Cxxm32} and
\eqref{eq:Cxxm2}--\eqref{eq:Czzm2}.
For $\sigma_{zz}$ with $p=-2$, for example,
$\mean{\sin^2\theta\int\dd\varepsilon\,
\tau_0E_0^2(\varepsilon+w)^{-2}e^{-(\varepsilon+w)/T}}
\simeq\tau_0E_0^2/(2t)$, giving $C_{zz}=T/t$.
The equivalent transport-function identities are
$\int_0^\infty E^{-2}\Phi_{zz}(E)\,\dd E=1/t$ and
$\int_0^\infty E^{-3/2}\Phi_{zz}(E)\,\dd E
=16/(3\pi\sqrt t)$.
These determine $\mathcal K_p$ in
Eqs.~\eqref{eq:Kzzm2} and \eqref{eq:Kzzm32}.

\section{Derivation of the Sommerfeld coefficients}
\label{app:sommerfeld}

For $G_{xx}(E)=\tau_{0}E_{0}^{-p}(E-2t)E^{p}$,
\begin{align}
\frac{G_{xx}''(\mu_{0})}{G_{xx}(\mu_{0})}&=\frac{p}{\mu_{0}^{2}}\left[(p-1)+\frac{2\mu_{0}}{\EF}\right],
\label{eq:B1}\\
\frac{G_{xx}''''(\mu_{0})}{G_{xx}(\mu_{0})}&=\frac{p(p-1)(p-2)}{\mu_{0}^{4}}\left[(p-3)+\frac{4\mu_{0}}{\EF}\right],
\label{eq:B2}
\end{align}
where we used $\mu_{0}-2t=\EF$. For general order,
\begin{equation}
\frac{G_{xx}^{(n)}(\mu_{0})}{G_{xx}(\mu_{0})}=\frac{(p+1)^{\underline{n}}\,\mu_{0}-2t\,p^{\underline{n}}}{\EF\,\mu_{0}^{n}},
\label{eq:B3}
\end{equation}
where $x^{\underline{n}}=x(x-1)\cdots(x-n+1)$ is the falling factorial. For $G_{zz}=\tau_{0}E_{0}^{-p}E^{p}$, $G_{zz}^{(2k)}/G_{zz}=p(p-1)\cdots(p-2k+1)/\mu_{0}^{2k}$. Inserting these into Eq.~\eqref{eq:sommerfeld} gives Eqs.~\eqref{eq:sommxx} and \eqref{eq:sommzz}. Corrections from the lower integration limit and from the shift of $\mu$ are of order $e^{-(\mu_{0}-4t)/T}$ and are treated in Sec.~\ref{sec:edge}.

\section{Corrections of order $t/T$ in the nondegenerate regime}
\label{app:tcorr}

In the nondegenerate regime the Boltzmann normalization follows from Eq.~\eqref{eq:muhighT}: $z=e^{\mu/T}\simeq(\EF/T)(1+2t/T)$, since $e^{2t/T}/I_{0}(2t/T)=1+2t/T+O(t^{2}/T^{2})$. For $\sigma_{xx}$ we need $F(w)\equiv\int_{0}^{\infty}\dd\varepsilon\,\varepsilon\,\tau(\varepsilon+w)e^{-(\varepsilon+w)/T}=\int_{w}^{\infty}(E-w)\tau(E)e^{-E/T}\dd E$, with $\partial F/\partial w=-\int_{w}^{\infty}\tau(E)e^{-E/T}\dd E$. Measuring energies in units of $T$ ($a=w/T$) and dropping the factor $\tau_{0}E_{0}^{-p}T^{p+2}$:
\begin{align}
p>-1:&\quad F=\Gamma(p+2)-a\,\Gamma(p+1)+O(a^{p+2}),
\label{eq:D1}\\
p=-1:&\quad F=e^{-a}-aE_{1}(a)\simeq1-a\left[1-\gamma-\ln a\right],
\label{eq:D2}\\
p=-\tfrac32:&\quad F\simeq\sqrt\pi-4\sqrt a+2\sqrt\pi\,a,
\label{eq:D3}\\
p=-2:&\quad F=(1+a)E_{1}(a)-e^{-a}\nonumber\\
&\quad\ \simeq-\ln a-\gamma-1+a\left(2-\gamma-\ln a\right).
\label{eq:D4}
\end{align}
Averaging over $\theta$ with $\mean{w}=2t$, $\mean{\sqrt w}=4\sqrt t/\pi$, $\mean{\ln w}=\ln t$, and $\mean{w\ln w}=2t\ln t+2t$, and multiplying by the normalization factor $1+2t/T$, gives Eqs.~\eqref{eq:tcorrxx}, \eqref{eq:tcorrxxm1}, \eqref{eq:Cxxm32}, and \eqref{eq:Cxxm2}. For $\sigma_{zz}$ we need $H(w)\equiv\int_{w}^{\infty}\tau(E)e^{-E/T}\dd E$, weighted by $2\sin^{2}\theta$; in the same units (dropping $\tau_{0}E_{0}^{-p}T^{p+1}$):
\begin{align}
p>0:&\quad H=\Gamma(p+1)-\frac{a^{p+1}}{p+1}+\cdots,
\label{eq:D5}\\
p=0:&\quad H=e^{-a}\simeq1-a,
\label{eq:D6}\\
p=-1:&\quad H=E_{1}(a)\simeq-\ln a-\gamma+a,
\label{eq:D7}\\
p=-\tfrac32:&\quad H\simeq2a^{-1/2}-2\sqrt\pi+2\sqrt a,
\label{eq:D8}\\
p=-2:&\quad H=\frac{e^{-a}}{a}-E_{1}(a)\simeq\frac1a-1+\ln a+\gamma.
\label{eq:D9}
\end{align}
With $2\mean{\sin^{2}\theta}=1$, $2\mean{\sin^{2}\theta\,w}=2t$, $2\mean{\sin^{2}\theta\ln w}=\ln t+\tfrac12$, $2\mean{\sin^{2}\theta\,w^{-1/2}}=8/(3\pi\sqrt t)$, and $2\mean{\sin^{2}\theta\,w^{-1}}=1/t$, and the normalization factor, these give Eqs.~\eqref{eq:tcorrzz}--\eqref{eq:Czzm2}. For $p>0$ the term $a^{p+1}$ in Eq.~\eqref{eq:D5} is of higher order than $t/T$, so that only the normalization contributes; for $p=0$ the two $O(t/T)$ contributions cancel.

\section{Lorenz ratio: moments and singular limits}
\label{app:WF}

Using the Fermi-window moments of Sec.~\ref{sec:WFdegenerate}, expansion of Eq.~\eqref{eq:Lorenz} through $T^{4}$ for $G(E)=E^{s}$, with $G^{(j)}(\mu)=s^{\underline{j}}\mu^{s-j}$, gives Eqs.~\eqref{eq:lambda2} and \eqref{eq:lambda4}. For $s=1$, $K_{0}\propto\mu$, $K_{1}\propto\pi^{2}T^{2}/3$, $K_{2}\propto\pi^{2}T^{2}\mu/3$ exactly (up to exponentially small terms). For in-plane $p=0$, $G_{xx}\propto E-2t$ instead gives $K_0\propto\EF$ and $K_2\propto\pi^2T^2\EF/3$, with $K_1$ unchanged, up to exponentially small terms. These moments yield Eq.~\eqref{eq:WFs1}.

In the nondegenerate regime the small-$t/T$ estimates use $M_n=\int\dd E\,E^{n+p}\Phi(-\partial f/\partial E)$, with $\tau_0E_0^{-p}$ suppressed, so that $\mathcal L/L_{0}=(3/\pi^{2})[M_{2}/M_{0}-(M_{1}/M_{0})^{2}]/T^{2}$. For $\sigma_{zz}$ with $p=-2$: $M_{0}\simeq f_{0}'/t$ with $f_{0}'\equiv(-\partial f/\partial E)_{E=0}=z/[T(1+z)^{2}]$; $M_{1}\simeq f_{0}'L_{zz}$ (the same logarithm as in $C_{zz}(-1)$); and $M_{2}=\int\Phi_{zz}(-\partial f/\partial E)\dd E\simeq f(0)=z/(1+z)$. Hence $M_2/M_0\simeq tT(1+z)$; the mean-square subtraction and same-order corrections are included in the remainder of Eq.~\eqref{eq:WFzzm2}. For $p=-3/2$: $M_{0}\simeq f_{0}'\cdot16/(3\pi\sqrt t)$ and $M_{2}\simeq\Gamma(\tfrac32)T^{1/2}\F_{1/2}(z)$, with $(M_{1}/M_{0})^{2}$ of higher order in $t/T$, giving Eq.~\eqref{eq:WFzzm32}. For the logarithmic cases, $M_{0}T=xL(1-\tfrac32x)+2\ln2\,x^{2}+O(x^{3})$ with $x=\EF/T$ (Sec.~\ref{sec:nondegenerate}), $M_{1}=f(0)=z/(1+z)\simeq x(1-x/2)$, and $M_{2}=T\ln(1+z)=\EF$, which give Eqs.~\eqref{eq:WFzzm1} and \eqref{eq:WFxxm2}. For $\sigma_{xx}$ with $p=-3/2$ the only $O[(t/T)^{1/2}]$ correction is in $M_{0}\simeq z\sqrt{\pi}\,T^{-1/2}[1-(16/\pi^{3/2})(t/T)^{1/2}]$, and $M_{2}/M_{0}-(M_{1}/M_{0})^{2}=(T^{2}/2)[1+\tfrac12\cdot(16/\pi^{3/2})(t/T)^{1/2}]$, which gives Eq.~\eqref{eq:WFxxm32}.

The two-window estimate~\eqref{eq:WFbipolar} follows from the law of total variance for a two-component weight: $\mathrm{Var}=\sum_{a}w_{a}\mathrm{Var}_{a}+\sum_{a}w_{a}(\bar E_{a}-\bar E)^{2}$, with the Fermi-window component (weight $1/(1+r)$, mean $\simeq\mu$) and the band-edge component (weight $r/(1+r)$, mean $\simeq0$). The edge variance is neglected and the mean separation is approximated by $\mu_0$.

\section{Scattering exponents and their assumptions}
\label{app:scattering}
For an isotropic parabolic band in $D$ dimensions, elastic Born transport obeys
\begin{equation}
\tau_{\rm tr}^{-1}(k)\propto k^{D-2}\int\dd\Omega\,
|U(2k\sin(\chi/2))|^2(1-\cos\chi).
\label{eq:scattering_scaling}
\end{equation}
Here $k=|\bm k|$ is the electron wave number, $\chi$ is the scattering angle, $q=2k\sin(\chi/2)$ is the transferred wave number, $U(q)$ is the Fourier-space scattering potential, and $\dd\Omega$ denotes the angular integration measure. A squared kernel $|U(q)|^2\propto q^{-\nu}$ gives $p=(\nu+2-D)/2$ when the angular integral converges. Table~\ref{tab:scattering} lists limiting assignments; screening and confinement need not yield a single power law across the superlattice miniband.

\begin{appendixtable}
\caption{Transport-time exponents, $\tau\propto E^p$, for isotropic parabolic bands under the stated assumptions. A dash denotes no universal assignment. Phonon entries omit the additional temperature-dependent prefactor.}
\label{tab:scattering}
\begin{ruledtabular}
\begin{tabular}{>{\raggedright\arraybackslash}p{0.29\textwidth}cc>{\raggedright\arraybackslash}p{0.48\textwidth}}
Mechanism & $p$ (3D) & $p$ (2D) & Assumptions and physical origin\\
\hline
Short-range defects, alloy disorder, neutral contact scatterers & $-1/2$ & $0$ & Born contact potential: constant matrix element, scattering rate proportional to the DOS~\cite{Lundstrom2000,LavasaniBulmashDasSarma2019}.\\[3pt]
Deformation-potential acoustic phonons & $-1/2$ & $0$ & Unscreened, quasielastic equipartition; fixed confinement. $|M|^2\propto q$ cancels $N_q\propto T/q$, giving a rate proportional to $T$ times the DOS~\cite{BardeenShockley1950,Price1981}.\\[3pt]
Coulomb impurities, strong screening & $-1/2$ & $0$ & Fixed $q_s\gg k$: effectively a contact potential~\cite{Lundstrom2000,AndoFowlerStern1982}.\\[3pt]
Coulomb impurities, weak screening & $3/2$ & $1$ & $k\gg q_s$. 3D: $U\propto(q^2+q_s^2)^{-1}$, with a Coulomb logarithm. 2D: $U\propto(q+q_s)^{-1}$, in-plane impurities, negligible thickness/setback~\cite{Brooks1955,DasSarmaHwang2004}.\\[3pt]
Remote charged impurities & --- & $3/2$ & Impurity sheet at setback $a$, $ka\gg1$, fixed screening: small-angle transport gives $\tau\propto k^3$~\cite{AndoFowlerStern1982}.\\[3pt]
Neutral impurities, Erginsoy model & $0$ & --- & Energy-independent rate~\cite{Erginsoy1950,Itoh1994}; distinct from a constant 3D cross section, which gives $\tau\propto1/v$ and $p=-1/2$.\\[3pt]
Interface roughness & --- & $0$ & White in-plane disorder and fixed subband coupling, with screening neglected~\cite{AndoFowlerStern1982}. A bulk isotropic contact-disorder analogue gives $p_{\rm 3D}=-1/2$, not a general interface result.\\[3pt]
Piezoelectric acoustic phonons & $1/2$ & $\nu/2$ & Unscreened, quasielastic equipartition: 3D squared kernel $\propto T/q^2$; 2D exponent depends on confinement and screening~\cite{Price1981}.\\[3pt]
Polar optical (Fr\"ohlich) phonons & $1/2$ & $\nu/2$ & Quasielastic, equipartition limit $E,T\gg\hbar\omega_{\rm LO}$: 3D squared kernel $\propto T/q^2$; 2D exponent depends on confinement and screening~\cite{Lundstrom2000,Price1981}.\\
\end{tabular}
\end{ruledtabular}
Here $q_s$ is the screening wave number, $M$ is the electron--phonon matrix element, $N_q$ is the phonon occupation number, and $\omega_{\rm LO}$ is the longitudinal optical phonon angular frequency. For the 2D phonon entries, the effective squared kernel after confinement averaging is $\propto q_\parallel^{-\nu}$; $\nu=0,1,2$ gives $p=0,1/2,1$, respectively. Optical phonon scattering outside the stated limit is inelastic.
\end{appendixtable}

The exponents $p=-1,-3/2,-2$ used here are phenomenological probes of enhanced low-energy lifetimes, rather than assignments to these mechanisms.

\section{Validation of the numerical results}
\label{app:validation}

The two-dimensional conductivity integrals over $(\theta,\varepsilon)$, Eqs.~\eqref{eq:sxx2d}--\eqref{eq:szz2d}, and the one-dimensional transport-function integrals, Eqs.~\eqref{eq:sxx1d}--\eqref{eq:szz1d}, were evaluated independently, with the chemical potential determined self-consistently from Eq.~\eqref{eq:density}. The two methods agree within $10^{-9}$ at $T/\EF=0.01,0.13,100$.

\subsection{Electrical conductivity}

Selected numerical conductivities are listed in Table~\ref{tab:numerics}.

\begin{appendixtable}
\caption{Numerical conductivities $\sigma_{xx}/\sigma_{xx}^{0}$ (upper block) and $\sigma_{zz}/\sigma_{zz}^{0}$ (lower block) at fixed density for $t=0.01\,\EF$. Rows marked ``analytic'' give the nondegenerate asymptotics, Eqs.~\eqref{eq:highTxx}--\eqref{eq:csinglog}, including the $t/T$ corrections and the quantum corrections to second order in $\EF/T$. At $T\le0.05\,\EF$ the analytical results of Secs.~\ref{sec:degenerate}--\ref{sec:edge} agree with the tabulated values in every digit shown (Table~\ref{tab:validation}).}
\label{tab:numerics}
\begin{ruledtabular}
\begin{tabular}{cccccccc}
$T/\EF$ & $p=-2$ & $p=-3/2$ & $p=-1$ & $p=0$ & $p=1$ & $p=3/2$ & $p=2$\\
\hline
\multicolumn{8}{c}{$\sigma_{xx}/\sigma_{xx}^{0}$}\\
$0.01$ & $0.9615$ & $0.9708$ & $0.9804$ & $1.0000$ & $1.0203$ & $1.0308$ & $1.0414$ \\
$0.05$ & $0.9687$ & $0.9734$ & $0.9802$ & $1.0000$ & $1.0282$ & $1.0457$ & $1.0654$ \\
$0.1$ & $0.9959$ & $0.9824$ & $0.9796$ & $1.0000$ & $1.0529$ & $1.0920$ & $1.1404$ \\
$0.2$ & $1.1521$ & $1.0139$ & $0.9696$ & $1.0000$ & $1.1483$ & $1.2714$ & $1.4360$ \\
$0.5$ & $1.4443$ & $0.9375$ & $0.8353$ & $1.0000$ & $1.6269$ & $2.2518$ & $3.2651$ \\
$1$ & $1.0768$ & $0.6501$ & $0.6106$ & $1.0000$ & $2.5749$ & $4.6336$ & $8.8769$ \\
$2$ & $0.5368$ & $0.3481$ & $0.3828$ & $1.0000$ & $4.5476$ & $11.110$ & $29.179$ \\
$5$ & $0.1471$ & $0.1184$ & $0.1783$ & $1.0000$ & $10.531$ & $39.731$ & $162.15$ \\
$10$ & $4.729\times10^{-2}$ & $4.706\times10^{-2}$ & $9.422\times10^{-2}$ & $1.0000$ & $20.526$ & $108.68$ & $623.80$ \\
$10$ (analytic) & $4.735\times10^{-2}$ & $0.046852$ & $9.415\times10^{-2}$ & $1$ & $20.526$ & $108.68$ & $623.80$ \\
$100$ & $7.537\times10^{-4}$ & $1.707\times10^{-3}$ & $9.935\times10^{-3}$ & $1.0000$ & $200.52$ & $3334.5$ & $6.0233\times10^{4}$ \\
$100$ (analytic) & $7.537\times10^{-4}$ & $1.7069\times10^{-3}$ & $9.935\times10^{-3}$ & $1$ & $200.52$ & $3334.5$ & $6.0234\times10^{4}$ \\
\hline
\multicolumn{8}{c}{$\sigma_{zz}/\sigma_{zz}^{0}$}\\
$0.01$ & $0.9621$ & $0.9713$ & $0.9807$ & $1.0000$ & $1.0200$ & $1.0303$ & $1.0407$ \\
$0.05$ & $0.9854$ & $0.9858$ & $0.9884$ & $1.0000$ & $1.0200$ & $1.0332$ & $1.0486$ \\
$0.1$ & $1.1302$ & $1.0493$ & $1.0179$ & $0.99995$ & $1.0200$ & $1.0425$ & $1.0733$ \\
$0.2$ & $4.8058$ & $1.7279$ & $1.2012$ & $0.9933$ & $1.0199$ & $1.0791$ & $1.1687$ \\
$0.5$ & $24.153$ & $4.2669$ & $1.5830$ & $0.8646$ & $1.0173$ & $1.2556$ & $1.6472$ \\
$1$ & $23.237$ & $3.7004$ & $1.1940$ & $0.6321$ & $1.0126$ & $1.5522$ & $2.5952$ \\
$2$ & $11.846$ & $1.8506$ & $0.5945$ & $0.3935$ & $1.0079$ & $2.0371$ & $4.5678$ \\
$5$ & $2.9499$ & $0.4667$ & $0.1614$ & $0.1813$ & $1.0036$ & $3.0716$ & $10.551$ \\
$10$ & $0.8575$ & $0.1377$ & $5.151\times10^{-2}$ & $9.516\times10^{-2}$ & $1.0019$ & $4.2738$ & $20.546$ \\
$10$ (analytic) & $0.8575$ & $0.1371$ & $0.051556$ & $9.517\times10^{-2}$ & $1.0020$ & $4.2741$ & $20.547$ \\
$100$ & $9.8442\times10^{-3}$ & $1.638\times10^{-3}$ & $8.028\times10^{-4}$ & $9.950\times10^{-3}$ & $1.0002$ & $13.316$ & $200.54$ \\
$100$ (analytic) & $9.844\times10^{-3}$ & $1.638\times10^{-3}$ & $8.028\times10^{-4}$ & $9.950\times10^{-3}$ & $1.0002$ & $13.316$ & $200.54$ \\
\end{tabular}
\end{ruledtabular}
\end{appendixtable}

The analytical results of Sec.~\ref{sec:electrical} provide checks of the numerical calculation in their respective limits. We compare the numerical conductivities with the following analytical predictions, all evaluated with $t=0.01\,\EF$ and with no adjustable parameters.

\emph{Degenerate regime} ($T/\EF=0.01$, $0.05$, $0.1$): the conductivities obtained by applying the Sommerfeld expansion~\eqref{eq:sommerfeld} through $T^6$, with coefficients~\eqref{eq:ak} and derivatives exact in $t$ [Eqs.~\eqref{eq:B1}--\eqref{eq:B3}], without additional exponential factors or band-edge terms.

\emph{Nondegenerate regime} ($T/\EF=10$, $100$): Eqs.~\eqref{eq:highTxx} and \eqref{eq:highTzz} with the coefficients $C(p)$ including the displayed miniband corrections, Eqs.~\eqref{eq:tcorrxx}--\eqref{eq:Czzm2}, and the quantum corrections to second order in $\EF/T$, Eqs.~\eqref{eq:cquantum}, \eqref{eq:csingpow}, and \eqref{eq:csinglog}.
The relative deviations $|\sigma^{\rm analytic}/\sigma^{\rm numerical}-1|$ are listed in Table~\ref{tab:validation} for all fourteen conductivity functions. At $T=0.05\,\EF$ the deviations of the nonterminating series are consistent with the first omitted Sommerfeld term, $a_{4}(T/\EF)^{8}$ times the corresponding derivative, which is largest for the most negative $p$. At $T=0.1\,\EF$ the deviations remain below $10^{-4}$ for all nonsingular cases, while for the singular cases ($4.3\%$ for $\sigma_{zz}$ with $p=-2$) the limitation is the asymptotic character of the Sommerfeld series and the omission of the band-edge contribution.

\begin{appendixtable}
\caption{Relative deviation $|\sigma^{\rm analytic}/\sigma^{\rm numerical}-1|$ between the analytical asymptotic formulas and the numerical results, for $t=0.01\,\EF$, in the degenerate ($T/\EF\le0.1$) and nondegenerate ($T/\EF\ge10$) regimes. The analytical expressions used are specified in Appendix~\ref{app:validation}. Entries ``$<10^{-12}$'' indicate deviations below the reporting threshold.}
\label{tab:validation}
\begin{ruledtabular}
\begin{tabular}{cccccccc}
$T/\EF$ & $p=-2$ & $p=-3/2$ & $p=-1$ & $p=0$ & $p=1$ & $p=3/2$ & $p=2$\\
\hline
\multicolumn{8}{c}{$\sigma_{xx}$}\\
$0.01$ & $5.8\times10^{-12}$ & $<10^{-12}$ & $<10^{-12}$ & $<10^{-12}$ & $<10^{-12}$ & $<10^{-12}$ & $<10^{-12}$ \\
$0.05$ & $2.9\times10^{-6}$ & $4.3\times10^{-7}$ & $7.4\times10^{-8}$ & $<10^{-12}$ & $2.2\times10^{-10}$ & $4.5\times10^{-9}$ & $3.1\times10^{-10}$ \\
$0.1$ & $7.7\times10^{-4}$ & $3.3\times10^{-5}$ & $6.6\times10^{-5}$ & $<10^{-12}$ & $9.6\times10^{-6}$ & $1.2\times10^{-5}$ & $1.3\times10^{-5}$ \\
$10$ & $1.2\times10^{-3}$ & $4.4\times10^{-3}$ & $7.0\times10^{-4}$ & $<10^{-12}$ & $2.6\times10^{-5}$ & $1.8\times10^{-5}$ & $8.6\times10^{-6}$ \\
$100$ & $8.8\times10^{-6}$ & $1.6\times10^{-4}$ & $1.2\times10^{-5}$ & $<10^{-12}$ & $2.6\times10^{-7}$ & $2.1\times10^{-7}$ & $1.6\times10^{-7}$ \\
\multicolumn{8}{c}{$\sigma_{zz}$}\\
$0.01$ & $6.2\times10^{-11}$ & $2.3\times10^{-11}$ & $6.9\times10^{-12}$ & $<10^{-12}$ & $<10^{-12}$ & $<10^{-12}$ & $<10^{-12}$ \\
$0.05$ & $3.8\times10^{-5}$ & $1.3\times10^{-5}$ & $3.6\times10^{-6}$ & $2.1\times10^{-9}$ & $4.7\times10^{-11}$ & $9.1\times10^{-9}$ & $2.2\times10^{-10}$ \\
$0.1$ & $4.3\times10^{-2}$ & $7.0\times10^{-3}$ & $1.1\times10^{-3}$ & $4.6\times10^{-5}$ & $9.6\times10^{-7}$ & $3.3\times10^{-6}$ & $9.4\times10^{-6}$ \\
$10$ & $3.1\times10^{-5}$ & $4.2\times10^{-3}$ & $8.3\times10^{-4}$ & $4.3\times10^{-5}$ & $9.8\times10^{-5}$ & $6.9\times10^{-5}$ & $4.8\times10^{-5}$ \\
$100$ & $1.1\times10^{-5}$ & $1.6\times10^{-4}$ & $4.6\times10^{-6}$ & $4.7\times10^{-8}$ & $1.0\times10^{-6}$ & $7.0\times10^{-7}$ & $4.9\times10^{-7}$ \\
\end{tabular}
\end{ruledtabular}
\end{appendixtable}

\begingroup
\looseness=-1
\widowpenalty=10000
In the nondegenerate regime, deviations at $T=100\,\EF$
are at most $1.2\times10^{-5}$ for cases with analytic
$t$ corrections, consistent with the first omitted terms,
$O[(\EF/T)^3]$ and $O(\EF t/T^2)$.
The two $p=-3/2$ cases have larger deviations
(about $1.6\times10^{-4}$): their $t$ corrections are
nonanalytic, $O[(t/T)^{1/2}]$, and the first omitted cross
term is $O[(\EF/T)(t/T)^{1/2}]$.
At $T=10\,\EF$, the same hierarchy holds, with larger
deviations ($10^{-5}$--$10^{-2}$).
The numerics reproduce the exact $p=0$ result
\eqref{eq:xxp0exact} to floating-point precision at all
temperatures, and the chemical potential agrees with
Eqs.~\eqref{eq:mulowT} and \eqref{eq:muhighT} in their
respective limits.
\par\endgroup

\subsection{Lorenz ratio}

Table~\ref{tab:WFvalid} compares numerical Lorenz ratios from
Eqs.~\eqref{eq:Kn}--\eqref{eq:Lorenz} with analytical
approximations through their relative deviations.
The degenerate comparisons use the finite-$t$ Sommerfeld
expansion through $T^4$, without a band-edge mixture;
the nondegenerate comparisons use
Eqs.~\eqref{eq:WFclassical}--\eqref{eq:WFxxm2}.
At $T=0.01\,\EF$, nonzero algebraic errors are
$10^{-10}$--$10^{-7}$, consistent with the omitted $T^6$
terms and largest for the most negative $s$.
For convergent cases,
$\mathcal L_{xx}(p)=\mathcal L_{zz}(p+1)$ as $t\to0$.
The singular cases reproduce the hopping dependence
at all three $t$ values in Fig.~\ref{fig:WFt}.

\begin{appendixtable}
\caption{Relative deviation $|\mathcal L^{\rm analytic}/\mathcal L^{\rm numerical}-1|$ of the Lorenz ratio, $t=0.01\,\EF$. Degenerate rows ($T/\EF\le0.1$): finite-$t$ Sommerfeld expansion through $T^4$, without an edge mixture. Nondegenerate rows ($T/\EF\ge10$): Eqs.~\eqref{eq:WFclassical}--\eqref{eq:WFxxm2}.}
\label{tab:WFvalid}
\begin{ruledtabular}
\begin{tabular}{cccccccc}
$T/\EF$ & $p=-2$ & $p=-3/2$ & $p=-1$ & $p=0$ & $p=1$ & $p=3/2$ & $p=2$\\
\hline
\multicolumn{8}{c}{$\mathcal{L}_{xx}$}\\
$0.01$ & $1\times10^{-8}$ & $3\times10^{-9}$ & $4\times10^{-10}$ & $<10^{-12}$ & $3\times10^{-10}$ & $9\times10^{-10}$ & $8\times10^{-10}$ \\
$0.05$ & $3\times10^{-4}$ & $6\times10^{-5}$ & $9\times10^{-6}$ & $2\times10^{-8}$ & $5\times10^{-6}$ & $1\times10^{-5}$ & $1\times10^{-5}$ \\
$0.1$ & $2\times10^{-2}$ & $5\times10^{-4}$ & $2\times10^{-3}$ & $2\times10^{-4}$ & $3\times10^{-4}$ & $8\times10^{-4}$ & $6\times10^{-4}$ \\
$10$ & $7\times10^{-3}$ & $2\times10^{-3}$ & $2\times10^{-3}$ & $8\times10^{-7}$ & $2\times10^{-5}$ & $3\times10^{-6}$ & $3\times10^{-5}$ \\
$100$ & $6\times10^{-6}$ & $1\times10^{-4}$ & $2\times10^{-4}$ & $8\times10^{-9}$ & $2\times10^{-7}$ & $2\times10^{-8}$ & $3\times10^{-7}$ \\
\multicolumn{8}{c}{$\mathcal{L}_{zz}$}\\
$0.01$ & $1\times10^{-7}$ & $4\times10^{-8}$ & $2\times10^{-8}$ & $<10^{-12}$ & $<10^{-12}$ & $1\times10^{-10}$ & $3\times10^{-10}$ \\
$0.05$ & $3\times10^{-3}$ & $1\times10^{-3}$ & $4\times10^{-4}$ & $3\times10^{-7}$ & $9\times10^{-9}$ & $2\times10^{-6}$ & $4\times10^{-6}$ \\
$0.1$ & $4\times10^{-1}$ & $1\times10^{-1}$ & $3\times10^{-2}$ & $2\times10^{-3}$ & $2\times10^{-4}$ & $2\times10^{-4}$ & $3\times10^{-4}$ \\
$10$ & $3\times10^{-2}$ & $4\times10^{-3}$ & $8\times10^{-3}$ & $5\times10^{-4}$ & $2\times10^{-5}$ & $3\times10^{-5}$ & $3\times10^{-5}$ \\
$100$ & $6\times10^{-3}$ & $1\times10^{-4}$ & $5\times10^{-5}$ & $6\times10^{-6}$ & $2\times10^{-7}$ & $3\times10^{-7}$ & $3\times10^{-7}$ \\
\end{tabular}
\end{ruledtabular}
\end{appendixtable}

\bibliography{references}

\end{document}